\documentclass[twocolumn,preprintnumbers,amsmath,amssymb,10pt]{revtex4}
\usepackage{amsmath}
\usepackage{extarrows}
\usepackage[ruled]{algorithm2e}
\usepackage{graphicx}% Include figure files
\usepackage{dcolumn}% Align table columns on decimal point
\usepackage{bm}% bold math
\usepackage[all]{xy}
\usepackage{indentfirst}
\usepackage{amsmath}
\usepackage{multirow}
\usepackage{mathrsfs}
\usepackage{bbm}
\usepackage{euscript}
\usepackage{amssymb}
\usepackage{extarrows}
\usepackage{bbm}
\usepackage[ruled]{algorithm2e}
\usepackage{graphicx}
\usepackage{color}
\usepackage{appendix}
\newtheorem{theorem}{Theorem}
\newtheorem{theorems}{Theorem S\!\!}
\newtheorem{definition}{Definition}

\newtheorem{lemma}{Lemma}

\begin{document}

\title{Strong entanglement distribution of quantum networks}

\author{Xue Yang, Yan-Han Yang, Ming-Xing Luo}

\affiliation{The School of Information Science and Technology, Southwest Jiaotong University, Chengdu 610031, China}

\begin{abstract}
Large-scale quantum networks have been employed to overcome practical constraints of transmissions and storage for single entangled systems. Our goal in this article is to explore the strong entanglement distribution of quantum networks. We firstly show any connected network consisting of generalized EPR states and GHZ states satisfies strong CKW monogamy inequality in terms of bipartite entanglement measure. This reveals interesting feature of high-dimensional entanglement with local tensor decomposition going beyond qubit entanglement. We then apply the new entanglement distribution relation in entangled networks for getting quantum max-flow min-cut theorem in terms of von Neumann entropy and R\'{e}nyi-$\alpha$ entropy. We finally classify entangled quantum networks by distinguishing network configurations under local unitary operations. These results provide new insights into characterizing quantum networks in quantum information processing.
\end{abstract}

\maketitle

\section{Introduction}

Entanglement as an intriguing phenomenon has been considered to be the heart of quantum mechanics. It provides a crucial resource for quantum information processing, including quantum teleportation \cite{Bennett1}, quantum dense coding \cite{Bennett2}, quantum secret sharing \cite{Hillery}, and quantum cryptography \cite{Gisin}. The study of entanglement and its distribution reveals fundamental insights into the nature of quantum correlations \cite{Ekert(1991),Horodecki(2009)}, the features of many-body systems \cite{Eisert(2010),Amico(2008)}, and the potential limitations for quantum-enhanced technologies \cite{Dowling(2003)}.

One interesting feature of entanglement is the impossibility of sharing entanglement freely in multiparty quantum systems. This without any classical counterpart is known as the monogamy of entanglement (MOE) \cite{Terhal(2004),V.Coffman}, that is, entanglement satisfy special constraints on how they can be distributed among multipartite systems. The monogamy inequality became synonymous due to Coffman, Kundu, and Wootters (CKW) \cite{V.Coffman} as
\begin{eqnarray}
{\cal Q}_{\textsf{A}|\textsf{B}\textsf{C}}(\rho_{\textsf{A}\textsf{B}\textsf{C}})\geq {\cal Q}_{\textsf{A}|\textsf{B}}(\rho_{\textsf{A}\textsf{B}})+{\cal Q}_{\textsf{A}|\textsf{C}}(\rho_{\textsf{A}\textsf{C}})
\label{eqn0}
\end{eqnarray}
with specific entanglement measure ${\cal Q}$. Here, $\rho_{\textsf{A}\textsf{B}}= \textrm{Tr}_{\textsf{C}}(\rho_{\textsf{A}\textsf{B}\textsf{C}})$ denotes the reduced state of parties $\textsf{A}$ and $\textsf{B}$, and analogously for $\rho_{\textsf{A}\textsf{C}}$. $\textsf{A}|\textsf{B}\textsf{C}$ is for the bipartite split. The first example is related the squared concurrence between bipartitions \cite{Terhal(2004),V.Coffman} for tripartite states and $N$-qubit states \cite{T.J.Osborne}. Some other monogamy inequalities are built for multiqubit systems in terms of the $\alpha$-th power of entanglement of formation (EOF) ($\alpha\geq \sqrt{2}$) \cite{Oliveira(2014),Bai3PRL,Guo} or the $\alpha$-th power of concurrence ($\alpha\geq 2$) \cite{Fei1,Jin}. Similar results hold for qubit systems with the $\alpha$($\alpha\geq 2$)-th powers of Tsallis entropy \cite{Luo(2016)}, R\'{e}nyi entropy \cite{Kim(2010)R}, and Unified entropy \cite{KimBarry(2011)}. However, most of these well-known entanglement measures \cite{T.J.Osborne,Oliveira(2014),Bai3PRL,Bai(2014),Guo,Luo(2016),Sara(2016),R(2015),Khan(2019)} fail to satisfy the monogamy relation (\ref{eqn0}) for qubit systems except for the squashed entanglement \cite{Koashi(2004)}. A natural problem is to explore which entangled states allow the CKW inequality. One intriguing example is from higher-dimensional quantum systems \cite{Ou(2007),Luo(2016),Luo2021}.

It is difficult to characterize general high-dimensional entangled states because of exponential parameters. One special scenario is from distributive constructions, that is, it allows local tensor structures further regarded as quantum networks \cite{LuoF}. This kind of multipartite resources are applicable for large-distance quantum communication \cite{Zuk,Duan}, quantum internet \cite{Kimble,Wehner}, quantum secret sharing \cite{Luo2021}, distributed quantum computing \cite{Cirac(1999)}, blind quantum computing \cite{Broadbent (2009)} and quantum sensing \cite{Gottesman(2012),P. (2014)}. This intrigues an interesting problem of universal monogamy relations for higher-dimensional systems from quantum networks.

Our motivation in this paper is to explore entangled quantum networks by using quantum  entropy toward quantum communication. We firstly prove the standard CWK monogamy (\ref{eqn0}) holds for generic entangled quantum networks in terms of EOF and R\'{e}nyi entropy. This shows new features of entanglement distribution for high-dimensional quantum states going beyond qubit entanglement \cite{Terhal(2004),V.Coffman,Fei1} or single entanglement \cite{Bai3PRL,R(2015)}. We then apply the proposed entanglement distribution to resolve the max-flow min-cut problem for entangled networks even it fails with single multipartite entanglement \cite{Pira(2019)}. The new Max-flow Min-cut theorem holds for any quantum networks consisting of EPR states and GHZ states. Moreover, the entanglement entropy implies one way for classifying the configuration for entangled networks under local unitary operations. The present method can be adapted for
cyclic networks going beyond recent results \cite{Kela(2020),Aberg(2020)}. These results are interesting in entanglement theory, quantum communication and quantum information processing.

\section{Entanglement monogamy relation of quantum networks}

\begin{figure}
\begin{center}
\resizebox{240pt}{150pt}{\includegraphics{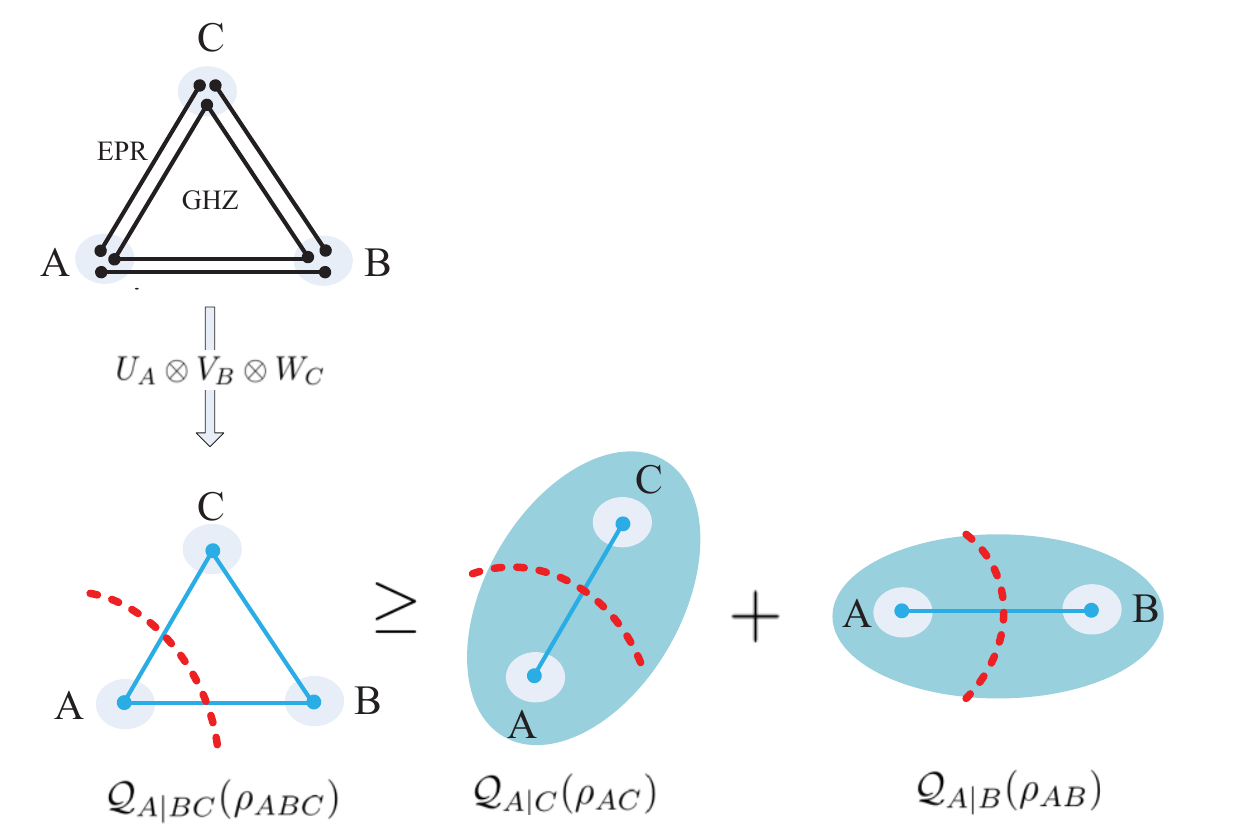}}
\end{center}
\caption{\small (Color online)  A new monogamy relation of tripartite entanglement inspired by tripartite quantum network consisting of one GHZ state and three EPR states. $U, V$ and $W$ are local unitary operations. ${\cal Q}$ is entanglement measure defined in the inequality (\ref{eqn0}).}
\label{trade-off}
\end{figure}

In comparison to single entanglement, an $n$-partite quantum network consists of various entangled states, as shown in Fig.\ref{trade-off}. Here, the local unitary operations may entangle these independent states into a new entanglement in global version. This generates high-dimensional quantum entangled states that are important resources in quantum information processing. Characterizing the general high-dimensional quantum systems is difficult because of exponential number of parameters. Our motivation here is to feature quantum networks vias entanglement distribution by using quantum entropy. The high-dimensional entanglement inspired by quantum networks allows local tensor decomposition that provides the possibility for evaluating new monogamy relations based on von Neumann entropy \cite{Nielsen} and  R\'{e}nyi entropy \cite{HHH(1996)}, Tsallis entropy \cite{Tsallis(1988)}, and Unified entropy \cite{Rathie(1991)}.

Consider an $n$-partite entangled quantum network $\mathcal{N}_q(\cal{A},\xi)$, where $\cal{A}$ denotes parties $\textsf{A}_1, \cdots, \textsf{A}_n$, and $\xi$ denotes entangled states. Assume any two parties $\textsf{A}_i$ and $\textsf{A}_j$ share the states $\varrho_{ij}(\theta_1), \cdots, \varrho_{ij}(\theta_s)$, $\sigma_{ij}(\varphi_1), \cdots, \sigma_{ij}(\varphi_t)$, and $\delta_{ij}(\vartheta_1), \cdots, \delta_{ij}(\vartheta_k)$, where $\varrho(\theta)$ denotes the density matrix of generalized EPR state \cite{EPR}: $|\phi(\theta)\rangle=\cos\theta|00\rangle+\sin\theta|11\rangle$, and $\sigma(\varphi)$ is the density matrix of generalized GHZ state \cite{GHZ}: $|\phi(\varphi)\rangle=\cos\varphi|0\rangle^{\otimes m}+\sin\varphi|1\rangle^{\otimes m}$ with any integer $m\geq 3$, and $\delta({\vartheta})=\cos^2\vartheta|00\rangle\langle00|+\sin^2\vartheta|11\rangle\langle11|)$ is the reduced density matrix of any two subsystems obtained by tracing out other  subsystems in a generalized GHZ state. Here, each party can perform any local unitary operations to entangle the local systems. With this formation, the total system may be regarded as an $n$-partite entanglement in high-dimensional Hilbert spaces. Informally, we show that the entanglement distribution for quantum networks with local unitary operations satisfies the monogamy inequality (\ref{eqn0}). One example of tripartite quantum network is shown in Fig.\ref{trade-off}.

\begin{theorem}\label{general00}
The entanglement distribution of quantum network $\mathcal{N}_q(\cal{A},\xi)$ is given by
\begin{eqnarray}
{\cal Q}_{\textsf{A}_i|\overline{\textsf{A}_i}}(\rho_{\textsf{A}_1\cdots\textsf{A}_n})
\geq\sum^N_{j=1,j\neq i}{\cal Q}_{\textsf{A}_i|\textsf{A}_j}(\rho_{\textsf{A}_i\textsf{A}_j}),
\label{eqngeneral01}
\end{eqnarray}
where ${\cal Q}$ denotes entanglement measure of EOF or R\'{e}nyi-$\alpha$ entropy, ${\cal Q}_{\textsf{X}|\textsf{Y}}$ is an entanglement measure of a composite quantum system with respect to the bipartite cut between $\textsf{X}$ and $\textsf{Y}$, and  $\overline{\textsf{A}_i}$ denotes all parties except for $\textsf{A}_i$.   The equality holds if and only if all GHZ states being shared by two parties.
\end{theorem}

The proof of Theorem \ref{general00} is provided in Appendix A based on the additivity of quantum entropy. It also includes entanglement distribution of quantum networks consisting of general entanglement (see Appendix B) or in terms of Tsallis entropy and Unified entropy (see Appendix C). Theorem 1 indicates a mutually exclusive relation of quantum networks between any pair of parties $\textsf{A}_i$ and $\textsf{A}_{j}$, which goes beyond the single entanglement of GHZ states being ruled out the monogamy inequality (\ref{eqn0}). The more interesting part is the present result provides the first generic monogamous relation for high-dimensional entanglement beyond qubit states \cite{Terhal(2004),V.Coffman,T.J.Osborne,Fei1,Luo(2016),Kim(2010)R,KimBarry(2011)}. It is also stronger than the existence result of high-dimensional entanglement \cite{Luo2021}. Theorem 1 may provide a better-than single-entanglement performance in quantum tasks. One example is device-independent quantum key distribution \cite{Horodecki(2009)} where eavesdropper's attacks are basically limited by the monogamy of entanglement.

\begin{figure}[htb]
\begin{center}
\resizebox{200pt}{90pt}{\includegraphics{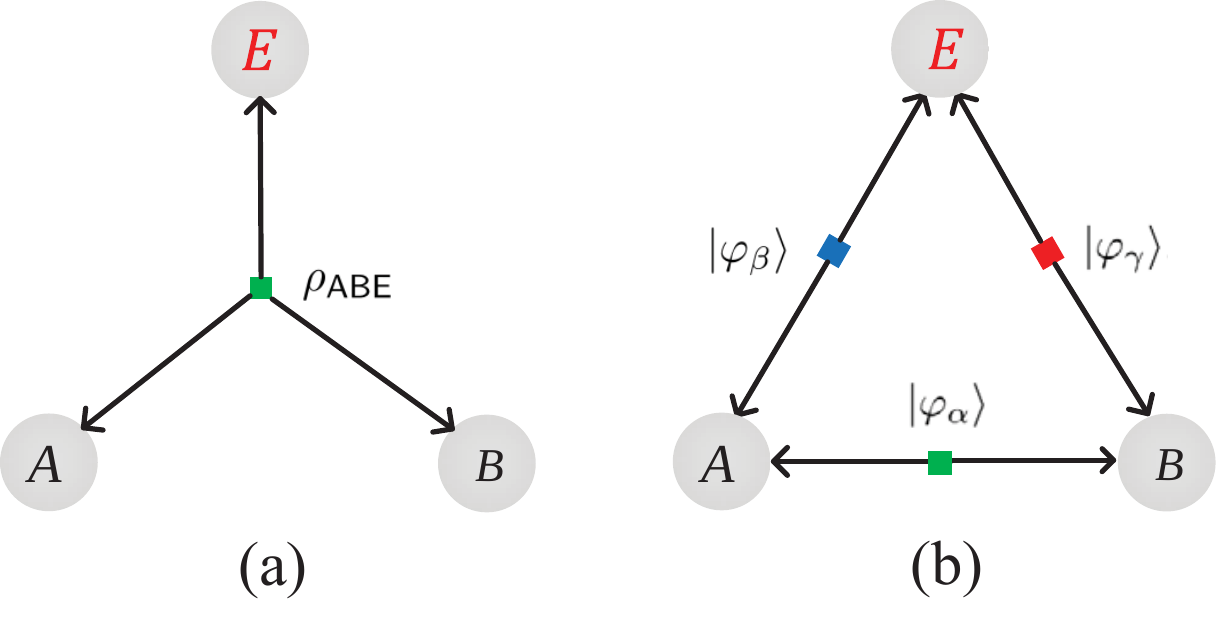}}
\end{center}
\caption{\small (Color online). Device-independent QKD with an attacker $E$. (a)  The standard attack model. (b) The new attack model based on triangle network.}
\label{QKD}
\end{figure}

\textbf{Example 1}. Consider a device-independent quantum key distribution model with entanglement $\rho_{\textsf{AB}}$ \cite{Ekert(1991)}. An eavesdropper Eve is represented by $\textsf{E}$ who may be correlated with $\rho_{\textsf{AB}}$. The total state is then denoted by $\rho_{\textsf{ABE}}$ satisfying $\textrm{Tr}_{\textsf{E}}(\rho_{\textsf{ABE}})=\rho_{\textsf{AB}}$, as shown in Fig.\ref{QKD}(a). Eve is assumed to control the source or measurement devices in a relaxed model. The entanglement between  Eve and Alice satisfies the monogamy relation \cite{Fei1}:
\begin{eqnarray}
 {\cal E}_{\textsf{A}|\textsf{E}}(\rho_{\textsf{AE}})\leq ({\cal E}^{\sqrt{2}}_{\textsf{A}|\textsf{BC}}(\rho_{\textsf{ABE}})- {\cal E}^{\sqrt{2}}_{\textsf{A}|\textsf{B}}(\rho_{\textsf{AB}}))^{\frac{1}{\sqrt{2}}},
\label{eqnconstraint1}
 \end{eqnarray}
where $ {\cal E}$ represents the EOF measure. The inequality (\ref{eqnconstraint1}) provides a monogamy relation for the information leakage in a device-independent model.

Instead, eavesdropper may hold local systems correlated with $\rho_{\textsf{AB}}$. That is, Eve fakes entangled sources $|\varphi_{\beta}\rangle$ and $|\varphi_{\gamma}\rangle$ between himself and respective Alice's device and Bob's device, as shown in Fig.\ref{QKD}(b). This generates a triangle quantum network which allows from Theorem 1 the new relation as
\begin{eqnarray}
 {\cal E}_{\textsf{A}|\textsf{E}}(\rho_{\textsf{AE}})\leq {\cal E}_{\textsf{A}|\textsf{BE}}(\rho_{\textsf{ABE}})- {\cal E}_{\textsf{A}|\textsf{B}}(\rho_{\textsf{AB}}).
 \label{eqnconstraint2}
\end{eqnarray}
Herein, the inequality (\ref{eqnconstraint2}) provides a stronger constraint for the information leakage in a device-independent model, which makes it harder for attacker to carry out the attack.

\section{Communication capacities of quantum networks}

Quantum networks provide new capabilities for generating and applying quantum entanglement. One fundamental problem is to determine the maximal rate achievable in end-to-end transmission of states. This has so far, been addressed for single entanglement \cite{Bennett1} or quantum networks \cite{Pira(2019)} consisting of bipartite entangled states. Our goal here is to investigate the capacities of general quantum networks ${\cal N}_q$ consisting of EPR states and GHZ states.

For any bipartite entanglement $|\varphi\rangle_{\textsf{A}\textsf{B}}$, it can be regarded as a quantum channel with the capacity \cite{Bennett1} $\mathcal{C}_{\textsf{A}\textsf{B}}=S(\rho_{\textsf{A}})=\mathcal{E}_{\textsf{A}|\textsf{B}}(|\varphi\rangle_{\textsf{A}\textsf{B}})$, where $S(\rho_{\textsf{A}})$ is the von Neumann entropy of the party $\textsf{A}$. As a result, the total capacity of quantum channel $\{\xi_{ij}\}$ connecting the party $\textsf{A}_i$ and $\textsf{A}_j$ in quantum network $\mathcal{N}_q$ is given by
\begin{eqnarray}
S(\rho^{j}_{i})&=&\mathcal{E}_{\textsf{A}_i|\textsf{A}_{j}}(\rho_{\textsf{A}_i\textsf{A}_j})
\nonumber
\\
&=&n_1+n_2,
 \label{eqnmarginal01}
\end{eqnarray}
where $n_1$ and $n_2$ denote the respective number of EPR states and GHZ states shared by party $\textsf{A}_i$ and $\textsf{A}_j$, $S(\rho^{j}_{i})$ represents the von Neumann entropy of the reduced density matrix  $\rho^j_i$ by tracing out the system of $j$, i.e., $\rho^j_i={\rm{Tr}}_j(|\xi_{ij}\rangle\langle\xi_{ij}|)$. This equality allows us to evaluate the capacity of quantum networks via von Neumann entropy. It is noteworthy that  R\'{e}nyi entropy is also available for evaluating channel capacity.

Before stating the main result, we explain the main idea inspired by classical network theory \cite{Voloshin(2009)}. Consider an $n$-partite entangled quantum network $\mathcal{N}_q({\cal{A}},\Omega)$. Let $\textbf{s}$ and $\textbf{t}$ be the source and sink, respectively. The capacity of one edge is a function $\mathcal{C}: \Omega \longrightarrow \mathbb{R}^{+}$, where the edge $(\textsf{A}_i,\textsf{A}_j)\in \Omega$ is connected by the channel $\xi_{ij}$ (consisting of EPR states and GHZ states) to transmit quantum information from the party $\textsf{A}_i$ to $\textsf{A}_j$. $\mathcal{C}(\xi_{ij})$ gives the maximum amount of flow through the channel $\xi_{ij}$. We also use the short-hand notation $\mathcal{C}(i,j)=\mathcal{C}(\xi_{ij})$. Similar to classical network \cite{Voloshin(2009)}, a \textit{flow} $f$ of quantum network $\mathcal{N}_q$ is an assignment of weights to channels satisfying: (i) Capacity conservation: $f_{i\rightarrow j}\leq \mathcal{C}(\xi_{ij})$ for any $\xi_{ij}\in \Omega$. (ii) Flow conservation: flow leaving from the party $j$ is equal to flow entering $j$ for $j\in \cal{A}\setminus\{\textbf{s},\textbf{t}\}$.
%The capacity of quantum network is further rewritten into an optimization
%\begin{eqnarray}
%\mathcal{C}=\max_{f}\mathcal{C}(\textbf{S}_1,\textbf{S}_2).
%\label{maxflow}
%\end{eqnarray}

To address the maximal flow of given network $\mathcal{N}_q({\cal{A}},\Omega)$, the main idea is to find the cut of associated undirected graph $\mathcal{G}=({\cal V},\Omega)$, where ${\cal V}$ consists of all vertices and $\Omega$ consists of all edges. Here, one vertex schematically denotes one party. One edge $(i,j)$ connected two vertices schematically denotes as an EPR state shared by the corresponding parties $\textsf{A}_i$ and $\textsf{A}_j$. The GHZ state shared by $m$ parties is associated with hyperedge connecting $m$ vertices. This correspondence is reasonable because of the equality $S(\rho_{\textsf{A}_i})=S(\rho_{\textsf{A}_j})$ for any GHZ state $\rho_{\textsf{A}_1\cdots \textsf{A}_m}$, $1\leq i<j\leq m$. Thus the quantum network $\mathcal{N}_q$ is associated with an undirected finite hypergraph $\mathcal{G}$. With this correspondence, the weight of each edge $(i,j)$ or hyperedge is unit, that is, the entanglement entropy of EPR state or GHZ state. A \textit{cut} $T_{cut}$ of $\mathcal{G}$ is a bipartition $(\textbf{S}_1,\textbf{S}_2)$ of ${\cal V}$ such that $\textbf{s}\in \textbf{S}_1$ and $\textbf{t}\in \textbf{S}_2$. Thus the capacity of cut equals to the sum of weights of edges leaving from $\textbf{S}_1$ to $\textbf{S}_2$, i.e., $\mathcal{C}(\textbf{S}_1,\textbf{S}_2)=\sum_{(i,j)| i\in \textbf{S}_1, j\in \textbf{S}_2}\mathcal{C}(i,j)$, where $\mathcal{C}(i,j)$ denotes the weight of edge $(i,j)$. Significantly, the flow in classical network obeys the celebrated Max-flow Min-cut Theorem \cite{Voloshin(2009)}. We prove similar result for the quantum settings.

\begin{theorem}\label{flowcut}
For a given quantum network $\mathcal{N}_q$ associated with undirected graph $\mathcal{G}$, the maximal flow $f_{\max}$ is equal to the minimal cut of $\mathcal{G}$, that is,
\begin{eqnarray}
f_{\max}=\min_{T_{cut}}\mathcal{C}(\textbf{S}_1,\textbf{S}_2).
 \label{eqnmax}
\end{eqnarray}
\end{theorem}

Theorem 2 states that the maximum flow from $\textbf{s}$ to $\textbf{t}$ in quantum network $\mathcal{N}_q$ equals to the minimum cut that separates $\textbf{s}$ and $\textbf{t}$ in the corresponding graph $\mathcal{G}$. This provides a general method similar to point-to-point quantum protocol \cite{Bennett1} for general quantum networks beyond recent result \cite{Pira(2019)}. The proof is shown in Appendix E. In applications, for a given quantum network $\mathcal{N}_q$ consisting of EPR states and GHZ states under local unitary operations, the network capacity can be evaluated by extending Ford-Fulkerson Algorithm \cite{Ford-Fulkerson} as follows.
\begin{algorithm}
\begin{itemize}
\item[Input:] An $n$-partite quantum network $\mathcal{N}_q$.
\item[Output:] A maximal flow.
\item{}Find the associated undirected graph $\mathcal{G}=({\cal V},\Omega)$.
\item{}Initialize $f(i,j)=0$ for all edges and hyperedges.
\item{}Find a path $p_{\textbf{s},\textbf{t}}$ from $\textbf{s}$ to $\textbf{t}$ with
$f(i,j)\leq \mathcal{C}(i,j)$ (augmenting path) on every edge and hyperedge.
\item{}Augment flow $f$ on edges or hyperedge along the path $p_{\textbf{s},\textbf{t}}$.
\item{}Repeat the procedure until there is no augmenting path from $\textbf{s}$ to $\textbf{t}$.
\end{itemize}
\end{algorithm}

\begin{figure}[htb]
\begin{center}
\resizebox{250pt}{110pt}{\includegraphics{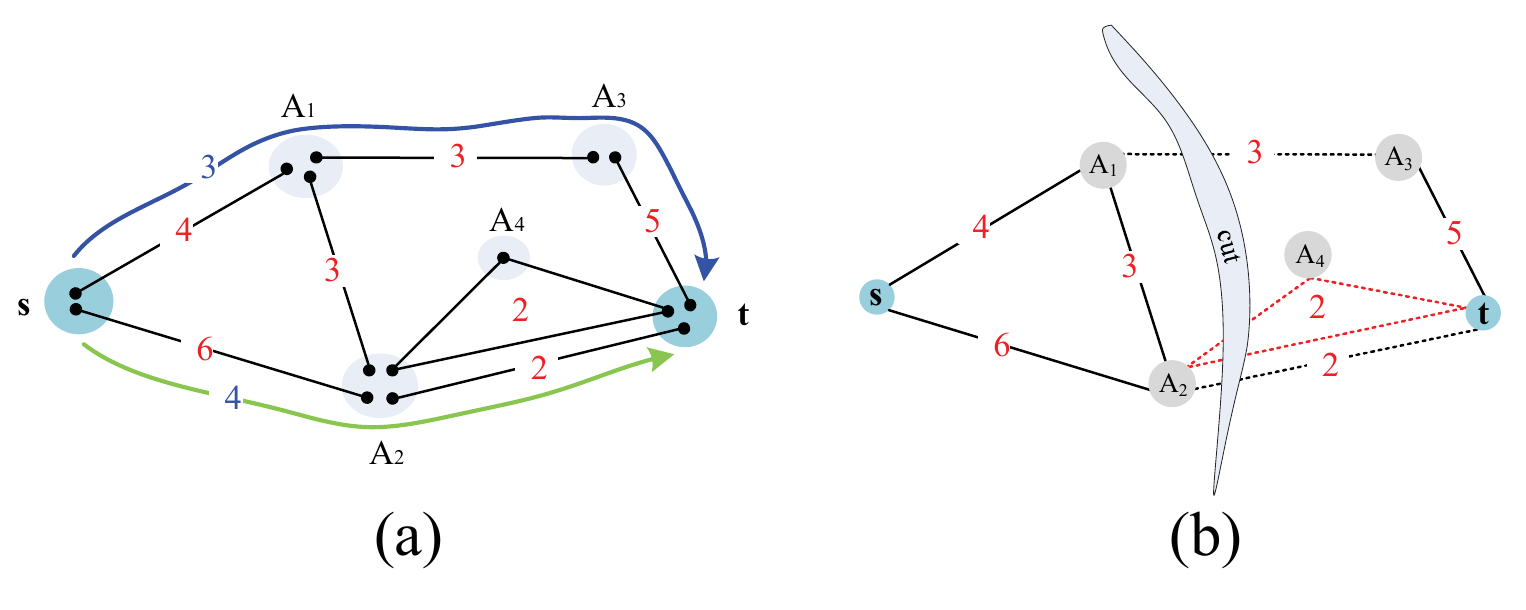}}
\end{center}
\caption{\small (Color online) (a) Max-flow of 6-partite quantum network ${\cal N}_q$. $\textbf{s}$ is source while $\textbf{t}$ is sink. Three parties $\textsf{A}_2,\textsf{A}_4$ and $\textbf{t}$ are connected by two GHZ states and the remaining parties are connected by multiple EPR states. The red integer on quantum channel denotes its capacity, i.e., the number of EPR states or GHZ states. The blue and green lines are two maximal flows. (b) Min-cut of the associated graph of ${\cal N}_q$. EPR state is denoted as one edge while GHZ state is denoted as one hyperedge (red line).  A source-sink cut achieves the min-cut of 7.}
\label{flowandcut}
\end{figure}

\textbf{Example 2}. Consider the maximization of the flow from $\textbf{s}$ to  $\textbf{t}$ as shown in Fig.\ref{flowandcut}(a) by applying to extended Ford-Fulkerson Algorithm. It is easy to check that the maximum flow from $\textbf{s}$ to $\textbf{t}$ is equal to 7, which is consistent with the minimum capacity of all source-sink cuts in a corresponding  undirected graph as shown in Fig.\ref{flowandcut}(b). The iterations of Example 2 are illustrated in Appendix \ref{iteration}.

\section{Network topology classification}

The actual quantum networks have a complex topology which significantly affects the communication efficiency. It is natural to consider how to identify network configuration for a given network. One method is to explore the measurement statistics by using the covariance matrix \cite{Kela(2020),Aberg(2020)} or the generalized Finner inequality \cite{Luo2021b} in a device-independent manner. The other is from the coherence theory \cite{Kraft(2021)}. These methods can be applied for characterizing quantum networks assisted by specific measurement. Our goal here is to consider a beyond problem of classifying configurations of given the set of networks with lower complexity. It is well-known that two local unitary equivalent multipartite entangled states have the same entanglement entropy with respect to any bipartition. We show that the converse  holds for quantum networks consisting of EPR states and GHZ states. Especially, for a given $n$-partite quantum network $\mathcal{N}_{q}$, denote the characteristic vector of quantum network $\mathcal{N}_q$ as
\begin{eqnarray}
S_{\mathcal{N}_q}=(S(\rho_{\mathsf{A}_1}), S(\rho_{\mathsf{A}_2}), \cdots, S(\rho_{\mathsf{A}_n})),
\label{vonv}
\end{eqnarray}
where $S(\rho_{\mathsf{A}_i})$ denotes the von Neumann entropy of the reduced state of the party $\mathsf{A}_i$ of $\mathcal{N}_{q}$. We present a configuration classification under the local unitary equivalence by virtue of the characteristic vector of quantum networks.

\begin{theorem}\label{equivalent}
Assume $\mathcal{N}_1$ and $\mathcal{N}_2$  are two quantum networks consisting of EPR states and GHZ state, where any two parties $\mathsf{A}_i$ and $\mathsf{A}_j$ in each network share no more than one entanglement. Then, $\mathcal{N}_1$ and $\mathcal{N}_2$ are unitary equivalence if and only if their characteristic vectors are equal to each other, that is, $S_{\mathcal{N}_1}=S_{\mathcal{N}_2}$.

\end{theorem}

\begin{figure}[htb]
\begin{center}
\resizebox{160pt}{140pt}{\includegraphics{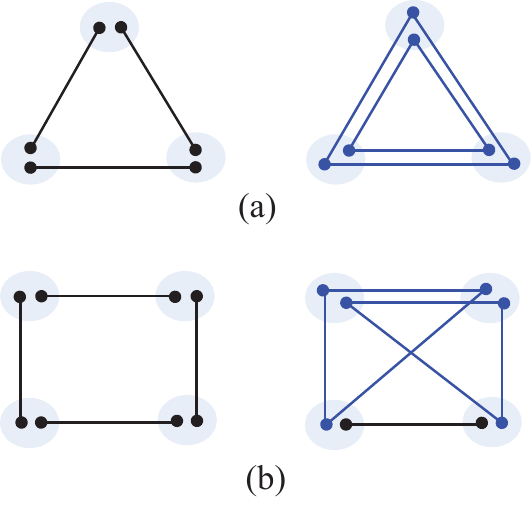}}
\end{center}
\caption{\small (Color online) Different networks with same characteristic vector. (a) Triangle network consisting of three EPR states (black lines) or two GHZ states (blue lines). (b) 4-partite network consisting of four EPR states or two GHZ states and one EPR states.}
\label{classify2}
\end{figure}

The proof of Theorem \ref{equivalent} is shown in  Appendix \ref{LUequivalence}.
Theorem \ref{equivalent} provides simple method to classify the given set of networks consisting of EPR states and GHZ states. In application, it only requires to evaluate $n$ number of von Neumann entropy $S(\rho_{\mathsf{A}_i})$, which has linear complexity beyond the exponential complexity for evaluating all entanglement entropies of $S(\rho_{\mathsf{A}_{i_1}\cdots \mathsf{A}_{i_t}})$, $1\leq i_1<\cdots<i_t\leq n$. The drawback is from the assumption of no more than one entanglement shared by given group of parties. Otherwise, two counterexamples are shown in Fig.\ref{classify2}. From Fig.\ref{classify2}(a), the triangle network consisting of three EPR states is inequivalent to its consisting of two GHZ states even for the same characteristic vector $(2,2,2)$. The other is the cyclic networks consisting four EPR states or two GHZ states and one EPR state in Fig.\ref{classify2}(b) with characteristic vector $(2,2,2,2)$. These can be extended for any cyclic quantum network in Appendix \ref{mutualinformation10}. Thus a new method beyond the characteristic vector of von Neumann entropy should be explored to verify these cyclic quantum networks. One possibility is from Shannon mutual information \cite{Shannon(1998)} derived from local projection measurements in Appendix \ref{mutualinformation10}.

\section{Discussions and conclusion}

Quantum entropy provides an easy way for characterizing entanglement distribution of specific quantum networks. This strong monogamy relations implies interesting applications in quantum communications and quantum network configurations. The present method is useful for generic settings of quantum network consisting of EPR states and GHZ states. These kind of networks are universal for measurement based quantum computation \cite{RB}. A natural problem is for general entangled networks. Specially, it should be useful for exploring quantum networks consisting of mixed states, which may intrigue new features going beyond pure systems. Another is for the max-flow problem on noisy networks including dephasing and erasure channels.

To sum up, we have proved the strong monogamy relation for quantum networks consisting of any bipartite entangled pure states and generalized GHZ states. These relations provide new feature of high dimensional entanglement beyond qubit states. This implies a celebrated max-flow min-cut theorem on quantum networks based on von Neumann entropy. Finally, we presented a new network classification by using Shannon or von Neumann entropies of the measurement statistics. This provides a device-independent manner for verifying quantum network configuration under local unitary operations. The present  results are interesting in quantum entanglement, quantum networks and quantum information processing.

\section*{Acknowledgements}

We thank the helps of Shaoming Fei, Yu Guo, Xiuyong Ding, Xiubo Chen. This work was supported by the National Natural Science Foundation of China (Nos.61772437,62172341), Sichuan Youth Science and Technique Foundation (No.2017JQ0048), Fundamental Research Funds for the Central Universities (No.2018GF07), and Shenzhen Institute for Quantum Science and Engineering.

%%%%%%%%%%%%%%%%%%%%%%%%%%%%%%%%%%%%%%%%%%%%%%%%%%%%%
%%%%%%%%%%%%%%%%%%%%%%%%%%%%%%%%%%%%%%%%%%%%%%%%%%%%%

\appendix

\section{Proof of Theorem 1}
\label{measures}

\subsection{Entanglement measure based on quantum entropy}

In this subsection, we recall necessary bipartite entanglement measures based on quantum entropies. As a entanglement measure characterizing bipartite entanglement through von Neumann entropy, the entanglement of formation (EOF) ${\cal{E}}$ for a pure state $|\phi\rangle_{\textsf{A}\textsf{B}}$ on Hilbert space ${\cal H}_{\textsf{A}}\otimes {\cal H}_{\textsf{B}}$ is defined as \cite{Bennett(1996)3824}:
\begin{eqnarray}
{\cal{E}}(|\phi\rangle_{\textsf{A}\textsf{B}})=S(\rho_{\textsf{A}})=-{\rm{Tr}}(\rho_{\textsf{A}}\log_2\rho_{\textsf{A}}),
\label{eqn03}
\end{eqnarray}
where $\rho_{\textsf{A}}={\rm Tr}_{\textsf{B}}(|\phi\rangle_{\textsf{A}\textsf{B}}\langle\phi|)$ denotes the density operator of the subsystem $\textsf{A}$ obtained by tracing out the subsystem $\textsf{B}$, and $S(\rho_{\textsf{A}})$ denotes the von Neumann entropy of the system $\textsf{A}$. Note that  von Neumann entropy of two  density operators $\rho$  and $\sigma$ satisfies the additivity\cite{Nielsen}:
\begin{eqnarray}
S(\rho\otimes\sigma)=S(\rho)+S(\sigma).
\label{eqnvon}
\end{eqnarray}
For a bipartite mixed state $\rho_{\textsf{A}\textsf{B}}$ on Hilbert space ${\cal H}_{\textsf{A}}\otimes {\cal H}_{\textsf{B}}$, EOF is given by
\begin{eqnarray}
{\cal{E}}(\rho_{\textsf{A}\textsf{B}})=\inf_{\{p_i,|\phi_i\rangle\}}\sum_ip_i{\cal{E}}(|\phi_i\rangle_{\textsf{A}\textsf{B}}),
\label{eqn030}
\end{eqnarray}
where the infimum takes over all possible pure-state decompositions of $\rho_{\textsf{A}\textsf{B}}=\sum_ip_i|\phi_i\rangle_{\textsf{A}\textsf{B}}\langle\phi_i|$ with $p_i\geq0$, $\sum_ip_i=1$, and $|\phi_i\rangle_{\textsf{A}\textsf{B}}\in {\cal H}_{\textsf{A}}\otimes {\cal H}_{\textsf{B}}$.

As a generalization of EOF, the R\'{e}nyi-$\alpha$ entanglement measure $\mathcal{R}_\alpha$ can be defined by using the R\'{e}nyi-$\alpha$ entropy as \cite{Kim(2010)R}:
\begin{eqnarray}
\mathcal{R}_\alpha(|\phi\rangle_{\textsf{A}\textsf{B}})=S_\alpha(\rho_{\textsf{A}})=\frac{1}{1-\alpha}\log_2 {\rm{Tr}} (\rho^q_{\textsf{A}}).
\label{eqnR1}
\end{eqnarray}
where $S_\alpha(\rho_{\textsf{A}})$ denotes the R\'{e}nyi-$\alpha$ entropy of the system $\textsf{A}$.   Similar to the von Neumann entropy, R\'{e}nyi-$\alpha$ entropy has the additivity   \cite{Dam(2002)}:
\begin{eqnarray}
S_\alpha(\rho\otimes\sigma)=S_\alpha(\rho)+S_\alpha(\sigma)
\label{eqnRe}
\end{eqnarray}
for $\alpha>0$ with $\alpha\neq 1$. For a bipartite mixed state $\rho_{\textsf{A}\textsf{B}}$,  R\'{e}nyi-$\alpha$ entanglement $\mathcal{R}_\alpha$ is defined via the convex-roof extension as \cite{Kim(2010)R}:
\begin{eqnarray}
\mathcal{R}_\alpha(\rho_{\textsf{A}\textsf{B}})=\inf_{\{p_i,|\phi_i\rangle\}}\sum_ip_i\mathcal{R}_\alpha(|\phi_i\rangle_{\textsf{A}\textsf{B}})
\label{eqnR2}
\end{eqnarray}
with the minimum taking over all possible decompositions of $\rho_{\textsf{A}\textsf{B}}$.

For a bipartite pure state $|\phi\rangle_{\textsf{A}\textsf{B}}$ on Hilbert space ${\cal H}_{\textsf{A}}\otimes {\cal H}_{\textsf{B}}$, The Tsallis-$q$ entanglement measure $\mathcal{T}^{(q)}$ derived from  Tsallis-$q$ entropy is defined as \cite{Kim(2010)T}:
\begin{eqnarray}
\mathcal{T}^{(q)}(|\phi\rangle_{\textsf{A}\textsf{B}})=S_q(\rho_{\textsf{A}})=\frac{1}{q-1}(1-{\rm{Tr}}(\rho_{\textsf{A}}^q)),
\label{eqnT01}
\end{eqnarray}
where $S_q(\rho_{\textsf{A}})$ denotes the Tsallis-$q$ entropy of the system $\textsf{A}$, $q>0$ and $q\neq 1$. $S_q(\rho)$ converges to the von Neumann entropy when $q\to1$, that is, $\lim_{q \rightarrow1}S_q(\rho)=-{\rm Tr}\rho\log_2(\rho)=S(\rho)$ for any bipartite operator density $\rho$. For a bipartite mixed state $\rho_{\textsf{A}\textsf{B}}$, Tsallis-$q$ entropy entanglement $\mathcal{T}^{(q)}$ is defined by
\begin{eqnarray}
\mathcal{T}^{(q)}(\rho_{\textsf{A}\textsf{B}})=\inf_{\{p_i,|\phi_i\rangle\}}\sum_ip_i\mathcal{T}_q(|\phi_i\rangle_{\textsf{A}\textsf{B}}),
\label{eqnT10}
\end{eqnarray}
where the infimum is taken over all possible pure-state decompositions of $\rho_{\textsf{A}\textsf{B}}$.

For $q>1$, the quantum Tsallis entropy satisfies the subadditivity
property \cite{Raggio(1995)}:
\begin{eqnarray}
S_q(\rho\otimes\sigma)\leq S_q(\rho)+S_q(\sigma),
\label{eqnT02}
\end{eqnarray}
while for $0<q<1$, it satisfies
\begin{eqnarray}
S_q(\rho\otimes\sigma)\geq S_q(\rho)+S_q(\sigma).
\label{eqnT020}
\end{eqnarray}

Different from all the stated entropies, there exists a generalized entropy, that is, Unified $(q, s)$-entropy \cite{Rathie(1991)}, which involves two real parameters $q$ and $s$:
\begin{eqnarray}
S_{q,s}(\rho)=\frac{1}{(1-q)s}({\rm{Tr}}(\rho^q)^s-1)
\label{}
\end{eqnarray}
where $q, s\geq0$ and $q\neq1$, $s\neq0$. This entropy includes R\'{e}nyi-$\alpha$ entropy \cite{HHH(1996)} and Tsallis-$q$ entropy \cite{Tsallis(1988)} as special cases of $s\rightarrow0$ and $s\rightarrow1$, respectively. Moreover, for $q\to 1$, it converges to the von Neumann entropy \cite{Nielsen}. Using the Unified $(q, s)$-entropy, the Unified $(q, s)$ entanglement measure is defined for a bipartite pure state $|\phi\rangle_{AB}$ on Hilbert space ${\cal H}_{\textsf{A}}\otimes {\cal H}_{\textsf{B}}$ by \cite{KimBarry(2011)}:
\begin{eqnarray}
\mathcal{U}^{(q,s)}(|\phi\rangle_{\textsf{A}\textsf{B}})=S_{q,s}(\rho_{\textsf{A}})
\label{eqnU01}
\end{eqnarray}
for each $q, s\geq0$. For a bipartite mixed state $\rho_{\textsf{A}\textsf{B}}$, its entanglement measure is given by
\begin{eqnarray}
\mathcal{U}^{(q,s)}(\rho_{\textsf{A}\textsf{B}})=\inf_{\{p_i,|\phi_i\rangle\}}\sum_ip_i\mathcal{U}^{(q,s)}(|\phi_i\rangle_{\textsf{A}\textsf{B}}),
\label{eqnU10}
\end{eqnarray}
where the infimum is taken over all possible pure state decompositions of $\rho_{\textsf{A}\textsf{B}}$.

For $0<q<1$ and $s<0$, or $q\geq1$ and $s\geq0$, the quantum Unified $(q, s)$-entropy satisfies the subadditivity property \cite{Hu(2006)}:
\begin{eqnarray}
S_{q,s}(\rho\otimes\sigma)\leq S_{q,s}(\rho)+S_{q,s}(\sigma),
\label{eqnU02}
\end{eqnarray}
while for $q>1$ and $s<0$, or $0<q<1$ and $s>0$ it satisfies
\begin{eqnarray}
S_{q,s}(\rho\otimes\sigma)\geq S_{q,s}(\rho)+S_{q,s}(\sigma).
\label{eqnU20}
\end{eqnarray}

From all the aforementioned bipartite entanglement measures, they are subadditive, that is,
\begin{eqnarray}
{\cal Q}(\rho\otimes\sigma)\leq {\cal Q}(\rho)+{\cal Q}(\sigma),
\label{additivity}
\end{eqnarray}
where ${\cal Q}$ can be any one of ${\cal E}, \mathcal{R}_\alpha, \mathcal{T}^{(q)}$ and $\mathcal{U}^{(q,s)}$.

\subsection{Proof of Theorem 1}

\label{gEPRGHZ}

\begin{figure}[htb]
\begin{center}
\resizebox{240pt}{190pt}{\includegraphics{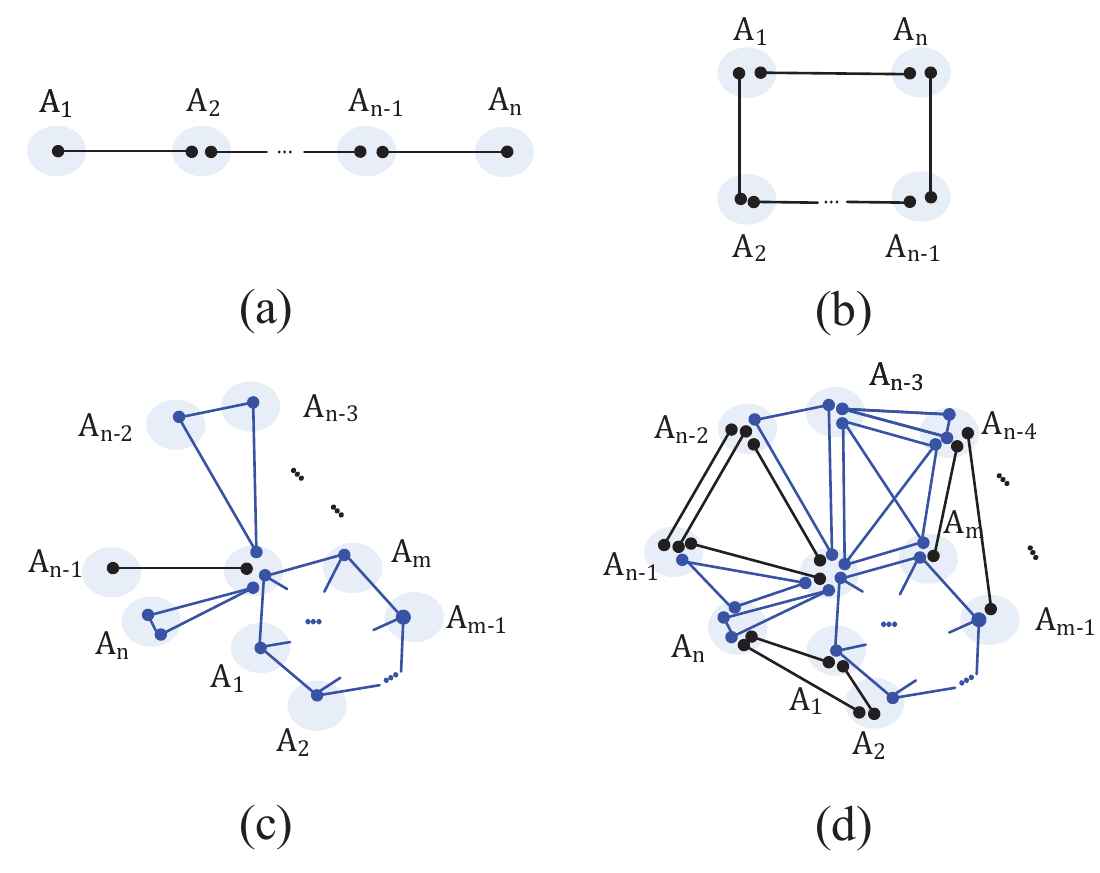}}
\end{center}
\caption{\small (Color online) Schematic quantum network consisting of generalized EPR states or generalized GHZ states. (a) A chain quantum network consisting of $n-1$ generalized EPR states. (b) A cyclic quantum network consisting of $n$ generalized EPR states. (c) A star quantum network consisting of generalized EPR states and generalized GHZ states. (d) A general quantum network consisting of generalized EPR states and generalized GHZ states.}
\label{eqnEPRGHZ}
\end{figure}

In this subsection, we prove Theorem 1 in terms of EOF and R\'{e}nyi-$\alpha$ entropy. The proof is completed by following the procedure from the chain network in case 1, cyclic network in case 2, star network in case 3 to general quantum network in case 4, as shown in Fig.\ref{eqnEPRGHZ}.

Consider a quantum network $\mathcal{N}_q(\cal{A},\xi)$, where $\cal{A}$ denotes parties $\textsf{A}_1, \cdots, \textsf{A}_n$, and $\xi$ denotes entangled states. Assume that any two parties $\textsf{A}_i$ and $\textsf{A}_j$ share the generalized EPR state \cite{EPR} as
\begin{eqnarray}
|\phi(\theta_s)\rangle=\cos\theta_s|00\rangle+\sin\theta_s|11\rangle,
\label{gEPR}
\end{eqnarray}
or generalized GHZ state \cite{GHZ} as
\begin{eqnarray}
|\phi(\varphi_t)\rangle=\cos\varphi_t|0\rangle^{\otimes m}+\sin\varphi_t|1\rangle^{\otimes m},
\end{eqnarray}
or only two qubits of one generalized GHZ state as
\begin{eqnarray}
\delta({\vartheta_k})=\cos^2\vartheta_k|00\rangle\langle00|+\sin^2\vartheta_k|11\rangle\langle11|.
\end{eqnarray}

\textbf{Case 1. Chain quantum networks}

From Fig.\ref{eqnEPRGHZ}(a), we consider a chain quantum network $\mathcal{N}^{(a)}_q$ consisting of $n$ parties $\mathsf{A}_1, \cdots, \mathsf{A}_n$, where each adjacent pair of $\mathsf{A}_i$ and $\mathsf{A}_{i+1}$ share $s_{i+1}$ generalized EPR states, i.e.,
\begin{eqnarray}
\xi_{i,i+1}=\mathop{\otimes}^{s_{i+1}}\limits_{s=1}|\phi({\theta_s})\rangle_{i,i+1}, 1\leq i\leq n-1,
 \end{eqnarray}
where $|\phi(\theta_s)\rangle$ is defined in Eq.(\ref{gEPR}). The total state $\rho_{\textsf{A}_1\cdots\textsf{A}_n}$ of $n$-partite system is given by
\begin{eqnarray}
 \rho_{\textsf{A}_1\cdots\textsf{A}_n}=\mathop{\otimes}^{n}_{i=1}\xi_{i,i+1}.
 \end{eqnarray}
It is easy to obtain
\begin{eqnarray}
{\cal Q}_{{\textsf{A}_i|\overline{\textsf{A}_i}}}(\rho_{\textsf{A}_1\cdots\textsf{A}_n})=\sum^n_{j=1,j\neq i}{\cal Q}_{{\textsf{A}_i|\textsf{A}_j}}(\rho_{\textsf{A}_i\textsf{A}_j}),
\label{eqnchain0}
\end{eqnarray}
where ${\cal Q}_{\textsf{A}_i|\overline{\textsf{A}_i}}$ and ${\cal Q}_{\textsf{A}_i|\textsf{A}_j}$ are given in Theorem \ref{general00}. The proofs are as follows. From Eq.(\ref{eqn03}), we have
\begin{eqnarray}
{\cal{E}}_{{\textsf{A}_i|\overline{\textsf{A}_i}}}(\rho_{\textsf{A}_1\cdots\textsf{A}_n})\nonumber&=&S(\rho_{\textsf{A}_i})
\\&=&\sum^{i+1}_{j=i-1,j\neq i}S(\mathop{\otimes}^{s_{j}}_{s=1}\rho_{i;j}(\theta_s))
 \label{eqnchain00}
\\&=&\sum^{i+1}_{j=i-1,j\neq i}\sum^{s_j}_{s=1}S(\rho_{i;j}(\theta_s)),
 \label{eqnchain10}
\end{eqnarray}
where ${\cal{E}}$ denotes the EOF. Eq.(\ref{eqnchain00}) is derived from $\rho_{\textsf{A}_i}=\mathop{\otimes}^{s_{j}}_{s=1}\rho_{i;j}(\theta_s)$, where $\rho_{i;j}({\theta_{s}})$ is the reduced density matrix of the systems owned by the party $\textsf{A}_i$ by tracing out all the subsystems in a generalized EPR state $|\phi({\theta_s})\rangle_{i,j}$ owned by $\textsf{A}_j$. Eq.(\ref{eqnchain10}) holds because of the additivity of the von Neumann entropy in Eq.(\ref{eqnvon}). On the other hand, the entanglement between the parties $\textsf{A}_i$ and $\textsf{A}_j$ is from entangled pairs associated with them. This implies that
\begin{eqnarray}
{\cal{E}}_{{\textsf{A}_i|\textsf{A}_{i-1}}}(\rho_{\textsf{A}_i\textsf{A}_{i-1}})
\nonumber&=&{\cal{E}}(\mathop{\otimes}^{s_{i-1}}_{s=1}|\phi({\theta_s})\rangle_{i,i-1})
\\&=&S(\mathop{\otimes}^{s_{i-1}}_{s=1}\rho_{i;i-1}(\theta_s))
\label{chain0}
\\&=&\sum^{s_{i-1}}_{s=1}S(\rho_{i;i-1}(\theta_s))
\label{chain1}
\end{eqnarray}
and
\begin{eqnarray}
{\cal{E}}_{{\textsf{A}_i|\textsf{A}_{i+1}}}(\rho_{\textsf{A}_i\textsf{A}_{i+1}})
\nonumber&=&{\cal{E}}(\mathop{\otimes}^{s_{i+1}}_{s=1}|\phi({\theta_s})\rangle_{i,i+1})
\\&=&S(\mathop{\otimes}^{s_{i+1}}_{s=1}\rho_{i;i+1}(\theta_s))
\label{chain2}
\\&=&\sum^{s_{i+1}}_{s=1}S(\rho_{i;i+1}(\theta_s)),
\label{chain3}
\end{eqnarray}
where Eqs.(\ref{chain0}) and (\ref{chain2}) from the definition of EOF in Eq.(A1). The additivity of the von Neumann entropy in Eq.(\ref{eqnvon}) yields to Eqs.(\ref{chain1}) and (\ref{chain3}).

Note that the party $\textsf{A}_i$ is only entangled with the adjacent parties $\textsf{A}_{i-1}$ and $\textsf{A}_{i+1}$. We have
\begin{eqnarray}
{\cal{E}}_{{\textsf{A}_i|\textsf{A}_{j}}}(\rho_{\textsf{A}_i\textsf{A}_j})=0,
j\not\in\{i-1, i, i+1\}.
 \label{eqnchain4}
\end{eqnarray}
Therefore, from Eqs.(\ref{chain1}) and (\ref{chain3}) it follows that
\begin{eqnarray}
\sum^n_{j=1}{\cal{E}}_{{\textsf{A}_i|\textsf{A}_{j}}}(\rho_{\textsf{A}_i\textsf{A}_{j}})\nonumber&=&\sum^{s_{i-1}}_{s=1}S(\rho_{i;i-1}(\theta_s))
\\&&+\sum^{s_{i+1}}_{s=1}S(\rho_{i;i+1}(\theta_s)).
 \label{eqnchain5}
\end{eqnarray}
Combining Eqs.(\ref{eqnchain10}) and (\ref{eqnchain5}), we conclude
\begin{eqnarray}
{\cal{E}}_{{\textsf{A}_i|\overline{\textsf{A}_i}}}(\rho_{\textsf{A}_1\cdots\textsf{A}_n})=\sum^n_{j=1,j\neq i}{\cal{E}}_{{\textsf{A}_i|\textsf{A}_j}}(\rho_{\textsf{A}_i\textsf{A}_{j}}).
\end{eqnarray}

Similarly, from the definition of R\'{e}nyi-$\alpha$ entropy entanglement in Eq.(\ref{eqnR1}) and the additivity in Eq.(\ref{eqnRe}), we can prove Eq.(\ref{eqnchain0}) holds for R\'{e}nyi-$\alpha$ entropy entanglement.

\textbf{Case 2. Cyclic quantum networks}

Consider a cyclic network $\mathcal{N}^{(b)}_q$ as shown in Fig.\ref{eqnEPRGHZ}(b), where  each pair of $\mathsf{A}_i$ and $\mathsf{A}_{i+1}$ share $s_{i+1}$ generalized EPR states, i.e., $\xi_{i,i+1}=\mathop{\otimes}^{s_{i+1}}\limits_{s=1}|\phi({\theta_s})\rangle_{i,i+1}$($1\leq i\leq n-1$). Moreover, the parties $\mathsf{A}_1$ and $\mathsf{A}_n$ share the quantum state $\xi_{1,n}=\mathop{\otimes}^{s_{n}}\limits_{s=1}|\phi({\theta_s})\rangle_{1,n}$. The total state $\rho_{\textsf{A}_1\cdots\textsf{A}_n}$ of $n$-partite system is denoted as
\begin{eqnarray}
 \rho_{\textsf{A}_1\cdots\textsf{A}_n}=\mathop{\otimes}^{n}\limits_{i=1}(\mathop{\otimes}^{s_n}\limits_{s=1}|\phi({\theta_s})
 \rangle_{1,n}\mathop{\otimes}^{s_{i+1}}\limits_{s=1}|\phi({\theta_s})\rangle_{i,i+1}).
 \end{eqnarray}
Similar to the proof of Eq.(\ref{eqnchain0}), it follows that
\begin{eqnarray}
{\cal Q}_{{\textsf{A}_i|\overline{\textsf{A}_i}}}(\rho_{\textsf{A}_1\cdots\textsf{A}_n})=\sum^n_{j=1,j\neq i}{\cal Q}_{{\textsf{A}_i|\textsf{A}_j}}(\rho_{\textsf{A}_i\textsf{A}_j}).
\label{eqncyclic}
\end{eqnarray}

\textbf{Case 3. Star quantum networks}

Consider an $n+1$-partite star quantum network $\mathcal{N}^{(c)}_q$ as shown in Fig. \ref{eqnEPRGHZ}(c). Suppose the party $\textsf{A}_0$ and any other party $\textsf{A}_j$ ($j=1, 2, \cdots, n$) share the quantum state $\xi_{0j}$ given by
\begin{eqnarray}
\xi_{0j}=\mathop{\otimes}^{s_j}_{s=1}\varrho_{0j}(\theta_s) \mathop{\otimes}^{t_j}_{t=1} \sigma_{0j}(\varphi_t)\mathop{\otimes}^{k_j}_{k=1}\delta_{0j}(\vartheta_k),
\label{starchannel}
\end{eqnarray}
where $\varrho({\theta_s})$ is the reduced density matrix of generalized EPR state \cite{EPR}: $|\phi({\theta_s})\rangle=\cos\theta_s|00\rangle+\sin\theta_s|11\rangle$, $\sigma({\varphi_t})$ is the reduced density matrix of generalized GHZ state \cite{GHZ}: $|\phi(\varphi_t)\rangle=\cos\varphi_t|0\rangle^{\otimes m}+\sin\varphi_t|1\rangle^{\otimes m}$, and $\delta({\vartheta_k})=\cos^2\vartheta_{k}|00\rangle\langle00|+\sin^2\vartheta_{k}|11\rangle\langle11|)$ is the reduced density matrix of any two subsystems by tracing out the remaining subsystems in a multipartite generalized GHZ state. Herein, $s_j$, $t_j$, and $k_j$ denote the numbers of $\varrho({\theta_s})$, $\sigma({\varphi_t})$, and $\delta({\vartheta_k})$, are shared by the party $\textsf{A}_0$ and any other party $\textsf{A}_j$, respectively.

The total state $\rho_{\textsf{A}_0\cdots\textsf{A}_n}$ of $n+1$-partite system is given by
\begin{eqnarray}
\rho_{\textsf{A}_0\cdots\textsf{A}_n}=\mathop{\otimes}^{n}_{j=1}
(\mathop{\otimes}^{s_j}_{s=1}\varrho_{0j}(\theta_s) \mathop{\otimes}^{t_j}_{t=1} \sigma_{0j}(\varphi_t)\mathop{\otimes}^{k_j}_{k=1}\delta_{0j}(\vartheta_k)).
\end{eqnarray}
The Theorem \ref{general00} means the following inequality
\begin{eqnarray}
{\cal Q}_{{\textsf{A}_i|\overline{\textsf{A}_i}}}(\rho_{\textsf{A}_0\cdots\textsf{A}_n})\geq\sum^n_{j=1}{\cal Q}_{{\textsf{A}_0|\textsf{A}_j}}(\rho_{\textsf{A}_0\textsf{A}_j}).
\label{star0}
\end{eqnarray}
where ${\cal Q}_{\textsf{A}_i|\overline{\textsf{A}_i}}$  and ${\cal Q}_{\textsf{A}_i|\textsf{A}_j}$ are given in Theorem \ref{general00}. The proof is as follows. From Eq.(\ref{eqn03}), we get
\begin{eqnarray}
&& {\cal{E}}_{\textsf{A}_0|\overline{\textsf{A}_0}}(\rho_{\textsf{A}_0\cdots\textsf{A}_n})
\nonumber
\\
&=&S(\rho_{\textsf{A}_0})
\nonumber
\\
&=&S(\mathop{\otimes}^{n}_{j=1}(\mathop{\otimes}^{s_j}_{s=1}\rho_{0;j}(\theta_s)\mathop{\otimes}^{t_j}_{t=1} \rho_{0;j}(\varphi_t)
\mathop{\otimes}^{k_{j}}_{k=1}\rho_{0;j}(\vartheta_k)))
\nonumber
\\
\nonumber&=&\sum^{n}_{j=1}(\sum^{s_j}_{s=1}S(\rho_{0;j}(\theta_s))+\sum^{t_j}_{t=1}
S(\rho_{0;j}(\varphi_t))
\\
&&+\sum^{k_j}_{k=1}S(\rho_{0;j}(\vartheta_k))),
\label{eqnstar1}
\end{eqnarray}
where Eq.(\ref{eqnstar1}) follows from the additivity of the von Neumann entropy in Eq.(\ref{eqnvon}). $\rho_{0;j}({\theta_{s}})$ is the reduced density matrix of the subsystems owned by $\textsf{A}_0$, which is obtained by tracing out all the subsystems owned by $\textsf{A}_j$ in a generalized EPR state $|\phi({\theta_s})\rangle_{0,j}$. $\rho_{0;j}(\varphi_{t})$ is the reduced density operator of the subsystem owned by $\textsf{A}_0$ with respect to generalized GHZ state shared by two parties $\textsf{A}_0$ and $\textsf{A}_j$. $\rho_{0;j}(\vartheta_k)$ is the reduced density matrix of the subsystems owned by  $\textsf{A}_0$, which is obtained by tracing out the other subsystems in a generalised GHZ state shared by multipartite parties. On the other side, the entanglement of $\textsf{A}_0$ and $\textsf{A}_{j}$ is generated by generalized EPR state and GHZ states.  This implies that
 \begin{eqnarray}
{\cal{E}}_{\textsf{A}_0|\textsf{A}_{j}}(\rho_{\textsf{A}_0\textsf{A}_j})\nonumber
 &=&{\cal{E}}(\mathop{\otimes}^{s_j}_{s=1}\varrho_{0j}(\theta_s) \mathop{\otimes}^{t_j}_{t=1} \sigma_{0j}(\varphi_t)\mathop{\otimes}^{k_j}_{k=1}\delta_{0j}(\vartheta_k))
\\
&\leq &
 {\cal{E}}(\mathop{\otimes}^{s_j}_{s=1}\varrho_{0j}(\theta_s))+{\cal{E}}(\mathop{\otimes}^{t_j}_{t=1} \sigma_{0j}(\varphi_t))
\label{eqnstar20}
\\
&=&\sum^{s_j}_{s=1}S(\rho_{0;j}(\theta_s))
 +\sum^{t_j}_{t=1}S(\rho_{0;j}(\varphi_t))
\label{eqnstar21}
 \end{eqnarray}
Here, by using the additivity of the entanglement measure ${\cal{E}}$ in Eq.(\ref{additivity}) and ${\cal{E}}(\delta_{0j}(\vartheta_k))=0$ iteratively, we can get the inequality (\ref{eqnstar20}). Eq.(\ref{eqnstar21}) follows from the definition of EOF in Eq.(\ref{eqn03}). Therefore, we have
\begin{eqnarray}
\sum^{n}_{j=1}{\cal{E}}_{\textsf{A}_0|\textsf{A}_{j}}(\rho_{\textsf{A}_0\textsf{A}_j})
\nonumber&\leq&\sum^{n}_{j=1}(\sum^{s_j}_{s=1}S(\rho_{0;j}(\theta_s))
\\
&&+\sum^{t_j}_{t=1}S(\rho_{0;j}(\varphi_t))).
 \label{eqnstar300}
\end{eqnarray}
Combining Eqs.(\ref{eqnstar1}) and (\ref{eqnstar300}), we get
\begin{eqnarray}
{\cal{E}}_{{\textsf{A}_i|\overline{\textsf{A}_i}}}(\rho_{\textsf{A}_0\cdots\textsf{A}_n})\geq\sum^n_{j=1}{\cal{E}}_{{\textsf{A}_i|\textsf{A}_j}}(\rho_{\textsf{A}_0\textsf{A}_j}).
\label{eqnstar301}
\end{eqnarray}

Note that for $k_j=0$, i.e., all $m$-qubit GHZ states are shared by two parties, we have
\begin{eqnarray}
{\cal{E}}_{\textsf{A}_0|\overline{\textsf{A}_0}}(\rho_{\textsf{A}_0\cdots\textsf{A}_n})
\nonumber&=&\sum^{n}_{j=1}(\sum^{s_j}_{s=1}S(\rho_{0;j}(\theta_s))
\\
&&+\sum^{t_j}_{t=1}S(\rho_{0;j}(\varphi_t))).
 \label{eqnstar30}
 \end{eqnarray}
Moreover, we have
\begin{eqnarray}
{\cal{E}}_{\textsf{A}_0|\textsf{A}_{j}}(\rho_{\textsf{A}_0\textsf{A}_j})\nonumber&=&\sum^{n}_{j=1}(\sum^{s_j}_{s=1}S(\rho_{0;j}(\theta_s))
\\&&+\sum^{t_j}_{t=1}S(\rho_{0;j}(\varphi_t))).
 \label{eqnstar31}
 \end{eqnarray}
So, combining Eqs.(\ref{eqnstar30}) and (\ref{eqnstar31}), we get
\begin{eqnarray}
{\cal{E}}_{\textsf{A}_0|\overline{\textsf{A}_0}}(\rho_{\textsf{A}_0\cdots\textsf{A}_n})=\sum^n_{j=1}{\cal{E}}_{\textsf{A}_0|\textsf{A}_{j}}(\rho_{\textsf{A}_0\textsf{A}_{j}}).
\label{eqnstar32}
 \end{eqnarray}

Similarly, the inequality (\ref{star0}) holds for R\'{e}nyi-$\alpha$ entropy entanglement defined in Eq.(\ref{eqnR1}) from the additivity of R\'{e}nyi-$\alpha$ entropy in Eq.(\ref{eqnRe}).

\textbf{Case 4. General quantum networks}

Now, we prove the monogamy for general network $\mathcal{N}_q(\cal{A},\xi)$ as shown in Fig.\ref{eqnEPRGHZ}(d), where any two parties $\textsf{A}_{i}$ and $\textsf{A}_{j}$ share the quantum state $\xi_{ij}$ given by
\begin{eqnarray}
\xi_{ij}=\mathop{\otimes}^{s_j}_{s=1}\varrho_{ij}(\theta_s) \mathop{\otimes}^{t_j}_{t=1} \sigma_{ij}({\varphi_t})\mathop{\otimes}^{k_j}_{k=1}\delta_{ij}(\vartheta_k),
\label{General0}
\end{eqnarray}
where $\varrho({\theta_s})$, $\sigma({\varphi_t})$, and $\delta({\vartheta_k})$ are defined in Eq.(\ref{starchannel}). $s_j$, $t_j$, and $k_j$ denote the numbers of $\varrho({\theta_s})$, $\sigma({\varphi_t})$, and $\delta({\vartheta_k})$ are shared by any two parties $\textsf{A}_{i}$ and $\textsf{A}_{j}$, respectively.

The total state $\rho_{\textsf{A}_1\cdots\textsf{A}_n}$ of $n$-partite system  is given by
\begin{eqnarray}
\rho_{\textsf{A}_1\cdots\textsf{A}_n}=\mathop{\otimes}^{n}_{\substack{i,j=1,\\i\neq j}}\xi_{ij},
\end{eqnarray}
where $\xi_{ij}$ has the form of Eq.(\ref{General0}).
Hence, Theorem \ref{general00} means that
\begin{eqnarray}
{\cal Q}_{{\textsf{A}_i|\overline{\textsf{A}_i}}}(\rho_{\textsf{A}_1\cdots\textsf{A}_n})\geq\sum^n_{j=1,j\neq i}{\cal Q}_{{\textsf{A}_i|\textsf{A}_j}}(\rho_{\textsf{A}_i\textsf{A}_j}),
\label{general0}
\end{eqnarray}
where ${\cal Q}_{\textsf{A}_i|\overline{\textsf{A}_i}}$  and ${\cal Q}_{\textsf{A}_i|\textsf{A}_j}$ are given in Theorem \ref{general00}.

The proof of the inequality (\ref{general0}) is as follows. Assume that any party $\mathsf{A}_i$ is entangled with $N$ parties of $\mathsf{A}_j$'s. From Eq.(\ref{eqn03}), we get
\begin{eqnarray}
{\cal{E}}_{{\textsf{A}_i|\overline{\textsf{A}_i}}}(\rho_{\textsf{A}_1\cdots\textsf{A}_n})
&=&
S(\rho_{\textsf{A}_i})
\\\nonumber
&=&S(\mathop{\otimes}^{N}_{j=1}(\mathop{\otimes}^{s_j}_{s=1}\rho_{i;j}(\theta_{s})\mathop{\otimes}^{t_j}_{t=1}\rho_{i;j}(\varphi_t)
\\
\nonumber&&\mathop{\otimes}^{k_{j}}_{k=1}\rho_{i;j}(\vartheta_k)))
\\
\nonumber&=&\sum^{N}_{j=1}(\sum^{s_j}_{s=1}S(\rho_{i;j}(\theta_{s}))+\sum^{t_j}_{t=1}S(\rho_{i;j}(\varphi_{t})))
\\
&&+\sum^{N}_{j=1}\sum^{k_j}_{k=1}S(\rho_{i;j}(\vartheta_k)),
\label{eqngeneric1}
\end{eqnarray}
where Eq.(\ref{eqngeneric1}) follows from the additivity of von Neumann entropy in Eq.(\ref{eqnvon}). Here, $\rho_{i;j}({\theta_{s}})$, $\rho_{i;j}(\varphi_{t})$, and $\rho_{i;j}(\vartheta_k)$ are defined in Eq.(\ref{eqnstar1}).

Note that the entanglement shared by the parties $\textsf{A}_i$ and $\textsf{A}_{j}$ is generated by generalized EPR states and GHZ states. It is sufficient to prove that
\begin{eqnarray}
{\cal{E}}_{{\textsf{A}_i|\textsf{A}_j}}(\rho_{\textsf{A}_i\textsf{A}_{j}})
&=&{\cal{E}}(\mathop{\otimes}^{s_j}_{s=1}\varrho_{ij}(\theta_s) \mathop{\otimes}^{t_j}_{t=1} \sigma_{ij}({\varphi_t})\mathop{\otimes}^{k_j}_{k=1}\delta_{ij}(\vartheta_k))
\nonumber
\\
&\leq &
{\cal{E}}(\varrho_{ij}(\theta_s))+{\cal{E}}(\mathop{\otimes}^{t_j}_{t=1} \sigma_{ij}({\varphi_t}))
\label{eqnstar22}
 \\
 &=&\sum^{s_j}_{s=1}S(\rho_{i;j}(\theta_s))+\sum^{t_j}_{t=1}S(\rho_{i;j}(\varphi_t)).
\label{eqnstar23}
\end{eqnarray}
From the additivity of the entanglement measure ${\cal{E}}$ in Eq.(\ref{additivity}) and ${\cal{E}}(\delta_{ij}(\vartheta_k))=0$, it is followed the inequality (\ref{eqnstar22}). Eq.(\ref{eqnstar23}) is derived from the definition of the entanglement measure ${\cal{E}}$  in Eq.(\ref{eqn03}). Combining Eqs. (\ref{eqngeneric1}) with (\ref{eqnstar23}), it yields to
\begin{eqnarray}
{\cal{E}}_{{\textsf{A}_i|\overline{\textsf{A}_i}}}(\rho_{\textsf{A}_1\cdots\textsf{A}_n})
&\geq& \sum^N_{j=1,j\neq i}{\cal{E}}_{{\textsf{A}_i|\textsf{A}_j}}(\rho_{\textsf{A}_i\textsf{A}_{j}})
\\&=&\sum^n_{j=1,j\neq i}{\cal{E}}_{{\textsf{A}_i|\textsf{A}_j}}(\rho_{\textsf{A}_i\textsf{A}_{j}}).
\label{eqngeneral}
\end{eqnarray}
Similar with the inequality (\ref{eqnstar301}), the equality (\ref{eqngeneral}) holds if and only if $k_j=0$, i.e., all the GHZ states are shared by two parties.

Similarly, the inequality (\ref{general0}) holds for the R\'{e}nyi-$\alpha$ entropy entanglement in Eq.(\ref{eqnR1}) from the additivity of R\'{e}nyi-$\alpha$ entropy in Eq.(\ref{eqnRe}). This has completed the proof.

\section{Generalizations of Theorem 1}
\label{egw}

In this section, we extend Theorem 1 to other networks.

\subsection{Quantum network with W states}

Consider a genuine tripartite entangled W state \cite{Dur}:
\begin{eqnarray}
|W\rangle=\frac{1}{\sqrt{3}}(|100\rangle+|010\rangle+|001\rangle).
\label{WW}
\end{eqnarray}
It is easy to obtain
\begin{eqnarray}
\nonumber&&{\cal{E}}(|W\rangle_{\textsf{A}|\textsf{B}_1\textsf{B}_2})=H(\frac{1}{3})\approx0.918,\\ &&{\cal{E}}(\rho_{\textsf{A}\textsf{B}_i})=H(\frac{1+\sqrt{1-C^2(\rho_{\textsf{A}\textsf{B}_i})}}{2})\approx0.55,
\label{eqnw0}
\end{eqnarray}
where ${\cal{E}}$ represents EOF in Eq.(\ref{eqn03}), $H(x)$ is binary
Shannon entropy given $H(x)=-x\log_2x-(1-x)\log_2(1-x)$ for $0\leq x \leq 1$ \cite{Nielsen}, and $C(\rho_{\textsf{A}\textsf{B}_i})=\frac{2}{3}$ ($i=1, 2$) is the concurrence \cite{Wootters(1998)} of two-qubit mixed state $\rho_{\textsf{A}\textsf{B}_i}$.

From Eq.(\ref{eqnw0}), we have
\begin{eqnarray}
{\cal{E}}(|W\rangle_{\textsf{A}|\textsf{B}_1\textsf{B}_2})\leq {\cal{E}}(\rho_{\textsf{A}\textsf{B}_1})+{\cal{E}}(\rho_{\textsf{A}\textsf{B}_2}),
 \label{}
\end{eqnarray}
that is, EOF ${\cal{E}}$ of W state is not monogamous. However, we prove that the monogamy relation of ${\cal{E}}$ holds for quantum networks consisting of GHZ states and W states. The monogamy of quantum network reaches a good saturation in the case that quantum network consists of $|W\rangle^{\otimes 5}\otimes |GHZ\rangle$, i.e., five pairs of W state and one GHZ state are shared by three parties, simultaneously. The main reason is as follows. For the GHZ state, we have
\begin{eqnarray}
 &&{\cal{E}}(|GHZ\rangle_{\textsf{A}|\textsf{B}\textsf{C}})=1,
\nonumber
\\
 &&
{\cal{E}}(\rho_{\textsf{A}\textsf{B}})={\cal{E}}(\rho_{\textsf{A}\textsf{C}})=0.
\label{eqnghz0}
\end{eqnarray}
From Eqs.(\ref{eqnw0}) and (\ref{eqnghz0}), we get the inequality
\begin{eqnarray}
\nonumber&&{\cal{E}}(|GHZ\rangle_{\textsf{A}|\textsf{B}\textsf{C}})+n\times {\cal{E}}(|W\rangle_{\textsf{A}|\textsf{B}_1\textsf{B}_2})\\&&\geq n{\cal{E}}(\rho_{\textsf{A}\textsf{B}_1})+ n{\cal{E}}(\rho_{\textsf{A}\textsf{B}_2})
\label{eqngw0}
\end{eqnarray}
for $n\leq5$. The inequality (\ref{eqngw0}) is optimal in the case of $n=5$, i.e., \begin{eqnarray}
1+5H(\frac{1}{3})\geq 5{\cal{E}}(\rho_{\textsf{A}\textsf{B}_1})+5{\cal{E}}(\rho_{\textsf{A}\textsf{B}_2}).
\label{QNW5}
\end{eqnarray}

From the inequality (\ref{QNW5}), we extend Theorem 1 to the network consisting of
EPR states, GHZ states and W states. Consider an $n$-partite entangled quantum network $\mathcal{N}_{q_1}(\cal{A},\xi)$, where $\cal{A}$ denotes parties $\textsf{A}_1, \cdots, \textsf{A}_n$, and $\xi$ denotes entangled states. Assume any two parties $\textsf{A}_i$ and $\textsf{A}_j$ share the state $\xi_{ij}$ given by
\begin{eqnarray}
\xi_{ij}=\mathop{\otimes}^{s_j}_{s=1}\varrho^{s}_{ij} \mathop{\otimes}^{t_j}_{t=1} \sigma^{t}_{ij}\mathop{\otimes}^{k_j}_{k=1}\delta^{k}_{ij}\mathop{\otimes}^{l_j}_{l=1}\omega^{l}_{ij},
\end{eqnarray}
where $\varrho$ denotes the density matrix of EPR state \cite{EPR}: $|\Phi\rangle=\frac{1}{\sqrt{2}}(|00\rangle+|11\rangle)$. $\sigma$ is the density matrix of generalized GHZ state \cite{GHZ}: $|\Psi\rangle=\frac{1}{\sqrt{2}}(|000\rangle+|111\rangle)$. $\delta=\frac{1}{\sqrt{2}}(|00\rangle\langle00|+|11\rangle\langle11|)$ is the reduced density matrix of any two subsystems in a GHZ state, and $\omega=\frac{2}{3}|00\rangle\langle00|+\frac{1}{3}|11\rangle\langle11|$ is the reduced density matrix of any two subsystems in W state in Eq.(\ref{WW}). $s_j$, $t_j$, $k_j$, and $l_j$ denote the numbers of $\varrho$, $\sigma$, $\delta$, and $\omega$, respectively. All these entangled states are shared by any two parties $\textsf{A}_i$ and $\textsf{A}_j$. Moreover, $l_j$ and $k_j$ satisfy $l_j\leq 5k_j$. Thus, the total state $\rho_{\textsf{A}_1\cdots\textsf{A}_n}$ of $n$-partite system  is given by
\begin{eqnarray}
\rho_{\textsf{A}_1\cdots\textsf{A}_n}=\mathop{\otimes}^{n}_{\substack{i,j=1,\\i\neq j}}(\mathop{\otimes}^{s_j}_{s=1}\varrho^{s}_{ij} \mathop{\otimes}^{t_j}_{t=1} \sigma^{t}_{ij}\mathop{\otimes}^{k_j}_{k=1}\delta^{k}_{ij}\mathop{\otimes}^{l_j}_{l=1}\omega^{l}_{ij}).
\end{eqnarray}

\begin{theorems}
\label{QNW}
The entanglement distribution of quantum network $\mathcal{N}_{q_1}$ satisfies
\begin{eqnarray}
{\cal{E}}_{{\textsf{A}_i|\overline{\textsf{A}_i}}}(\rho_{\textsf{A}_1\cdots\textsf{A}_n})\geq\sum^n_{j=1,j\neq i}{\cal{E}}_{{\textsf{A}_i|\textsf{A}_j}}(\rho_{\textsf{A}_i\textsf{A}_{j}}),
\label{eqngeneral011}
\end{eqnarray}
where ${\cal{E}}_{\textsf{X}|\textsf{Y}}$ represents EOF of a bipartite systems $\textsf{X}$ and $\textsf{Y}$, and $\overline{\textsf{A}_i}$ denotes all parties except for $\textsf{A}_i$.

\end{theorems}

\textit{Proof.} Assume that any party $\mathsf{A}_i$ is entangled with $N$ parties of $\mathsf{A}_j$'s. The total numbers of EPR, GHZ and W states are $n_1$, $n_2$, $n_3$, respectively. Any two parties $\mathsf{A}_i$ and $\mathsf{A}_j$ share $s_j$ pairs of EPR states $\varrho$, $t_j$ pairs of GHZ states $\sigma$, $k_j$ pairs of $\delta$, and $l_j$ pairs of $\omega$. As a result, it is easy to obtain that $\sum^N_{j=1}s_j= n_1$, $\sum^N_{j=1}t_j=n'_2$, $\sum^N_{k=1}k_j=2(n_2-n'_2)$, and $\sum^N_{l=1}l_j=2n_3$, where $n'_2$ is the number of GHZ states shared by two parties. There are $n_2-n'_2$ pairs of GHZ states shared by three parties. From the inequality (\ref{QNW5}), the entanglement distribution holds for quantum network consisting of GHZ states and W states from $l_j\leq5k_j$. This yields the inequality $n_3\leq 5(n_2-n'_2)$. From Eq.(\ref{eqn03}), we get
\begin{eqnarray}
\!\!\!\!\!\!\!\!\!{\cal{E}}_{{\textsf{A}_i|\overline{\textsf{A}_i}}}(\rho_{\textsf{A}_1\cdots\textsf{A}_n})\nonumber&=&S(\rho_{\textsf{A}_i})
 \label{eqnqnw10}
\\\nonumber&=&S((\frac{\mathbbm{1}}{2})^{\otimes n_1}\otimes (\frac{\mathbbm{1}}{2})^{\otimes n_2}\otimes\omega^{\otimes n_3})
 \label{eqnqnw11}
\\&=&n_1S(\frac{\mathbbm{1}}{2})+n_2S(\frac{\mathbbm{1}}{2})
  +n_3S(\omega)
 \label{eqnqnw12}
\\&=&n_1+n_2+n_3H(\frac{1}{3}),
 \label{eqnqnw13}
\end{eqnarray}
where Eq.(\ref{eqnqnw12}) follows from the additivity of the von Neumann entropy in Eq.(\ref{eqnvon}). In the equality (\ref{eqnqnw13}), we have used $S(\frac{\mathbbm{1}}{2})=1$ and $S(\omega)=H(\frac{1}{3})$.

Moreover, we have
\begin{eqnarray}
\!\!\!\!\!\!{\cal{E}}_{{\textsf{A}_i|\textsf{A}_{j}}}(\rho_{\textsf{A}_i\textsf{A}_{j}})\nonumber&=&{\cal{E}}(\mathop{\otimes}^{s_j}_{s=1}\varrho^s_{ij} \mathop{\otimes}^{t_j}_{t=1} \sigma^t_{ij}\mathop{\otimes}^{k_j}_{k=1}\delta^{k}_{ij}\mathop{\otimes}^{l_j}_{l=1}\omega^{l}_{ij})
\\\nonumber&\leq&{\cal{E}}(\mathop{\otimes}^{s_j}_{s=1}\varrho^s_{ij})+{\cal{E}}(\mathop{\otimes}^{t_j}_{t=1}\sigma^t_{ij})
\\&&+{\cal{E}}(\mathop{\otimes}^{l_j}_{l=1}\omega^{l}_{ij})
\label{QNW00}
\\\nonumber&\leq&\sum^{s_j}_{s=1}S(\frac{\mathbbm{1}}{2})+\sum^{t_j}_{t=1}S(\frac{\mathbbm{1}}{2})
\\&&+\sum^{l_j}_{l=1}{\cal{E}}(\omega^{l}_{ij})
\label{QNW01}
\\&=&\sum^{s_j}_{s=1}1+\sum^{t_j}_{t=1}1+\sum^{l_j}_{l=1}{\cal{E}}(\omega)
\\&=&s_j+t_j+l_j{\cal E}(\omega),
 \label{QNW02}
\end{eqnarray}
where the inequality (\ref{QNW00}) follows from the additivity of EOF in Eq.(\ref{additivity}). The inequality (\ref{QNW01}) is from the definition of EOF in Eq.(\ref{eqn03}) and the additivity of EOF in Eq.(\ref{additivity}). We therefore have
\begin{eqnarray}
\!\!\!\!\!\!\sum^N_{j=1}{\cal{E}}_{{\textsf{A}_i|\textsf{A}_{j}}}(\rho_{\textsf{A}_i\textsf{A}_{j}})\nonumber&\leq& \sum^N_{j=1}(s_j+t_j+l_j{\cal{E}}(\omega))
\\&=& n_1+n'_2+2n_3{\cal{E}}(\omega)
 \label{QNW10}
\\&\leq&n_1+n'_2+\frac{n_3}{5}+n_3H(\frac{1}{3})
 \label{QNW11}
\\&\leq&n_1+n_2+n_3H(\frac{1}{3}),
 \label{QNW12}
\end{eqnarray}
where the equality (\ref{QNW10}) is derived from the equalities: $\sum^N_{j=1}s_j= n_1$, $\sum^N_{j=1}t_j= n'_2$, and $\sum^N_{j=1}l_j=2n_3$. The inequality (\ref{QNW11}) follows from the inequality $5\times2{\cal{E}}(\omega)\leq 1+5H(\frac{1}{3})$ which is from the inequality (\ref{QNW5}). The inequality (\ref{QNW12}) holds for the inequality $n_3\leq 5(n_2-n'_2)$. Finally, we obtain
\begin{eqnarray}
{\cal{E}}_{{\textsf{A}_i|\overline{\textsf{A}_i}}}(\rho_{\textsf{A}_1\cdots\textsf{A}_n})\nonumber&\geq& \sum^N_{j=1,j\neq i}{\cal{E}}_{{\textsf{A}_i|\textsf{A}_{j}}}(\rho_{\textsf{A}_i\textsf{A}_{j}})
\\&=&\sum^n_{j=1,j\neq i}{\cal{E}}_{{\textsf{A}_i|\textsf{A}_{j}}}(\rho_{\textsf{A}_i\textsf{A}_{j}})
\end{eqnarray}
This completes the proof of Theorem \textbf{S\ref{QNW}}.

\subsection{Quantum network with arbitrary bipartite entangled pure states}
\label{bipartitestates}

In this subsection, consider an $n$-partite entangled quantum network $\mathcal{N}_{q_2}(\cal{A},\xi)$, where any two parties $\textsf{A}_i$ and $\textsf{A}_j$ share $s_j$ pairs of arbitrary bipartite entangled pure states $|\varphi^{1}\rangle_{ij}, |\varphi^{2}\rangle_{ij},\cdots, |\varphi^{s}\rangle_{ij}$. The joint state shared by two parties $\textsf{A}_i$ and $\textsf{A}_j$ is denoted by $\mathop{\otimes}^{s_j}_{s=1}|\varphi^{s}\rangle_{ij}$. The total state  $\rho_{\textsf{A}_1\cdots\textsf{A}_n}$ of $\mathcal{N}_{q_2}(\cal{A},\xi)$ is given by
\begin{eqnarray}
 \rho_{\textsf{A}_1\cdots\textsf{A}_n}=\mathop{\otimes}^{n}_{\substack{i,j=1,\\i\neq j}}\mathop{\otimes}^{s_j}_{s=1}|\varphi^{s}\rangle_{ij}.
\end{eqnarray}
We prove strong monogamy relation for quantum network $\mathcal{N}_{q_2}$.

\begin{theorems}\label{bipartiteE}
The entanglement distribution of quantum network $\mathcal{N}_{q_2}$ satisfies
\begin{eqnarray}
{\cal Q}_{{\textsf{A}_i|\overline{\textsf{A}_i}}}(\rho_{\textsf{A}_1\cdots\textsf{A}_n})
=\sum^n_{j=1,j\neq i}{\cal Q}_{{\textsf{A}_i|\textsf{A}_j}}(\rho_{\textsf{A}_i\textsf{A}_{j}}),
\label{eqnbipartite}
\end{eqnarray}
where ${\cal Q}_{\textsf{A}_i|\overline{\textsf{A}_i}}$  and ${\cal Q}_{\textsf{A}_i|\textsf{A}_j}$s are defined in Theorem \ref{general00}.

\end{theorems}

\emph{Proof.} Assume that there exist $N$ parties $\textsf{A}_j$'s who are entangled with  the party $\textsf{A}_i$. Note that any two parties $\textsf{A}_i$ and $\textsf{A}_j$ share the state of $\mathop{\otimes}^{s_j}_{s=1}|\varphi^{s}\rangle_{ij}$. It is sufficient to prove that
\begin{eqnarray}
{\cal{E}}_{{\textsf{A}_i|\overline{\textsf{A}_i}}}(\rho_{\textsf{A}_1\cdots\textsf{A}_n})\nonumber&=&{\cal{E}}(\mathop{\otimes}^{N}_{j=1}\mathop{\otimes}^{s_j}_{s=1}|\varphi^{s}\rangle_{ij})
\\\nonumber&=&S(\rho^j_i)
\\&=&S(\mathop{\otimes}^{N}_{j=1}\mathop{\otimes}^{s_j}_{s=1}\rho^{s}_{i;j})
\label{eqnbipartite0}
\\&=&\sum^{N}_{j=1}\sum^{s_j}_{s=1}S(\rho^{s}_{i;j}),
\label{eqnbipartite1}
\end{eqnarray}
where the equality (\ref{eqnbipartite0}) is due to the systems $\rho^{s}_{i;j}$'s are uncorrelated, that is,  $\rho^j_i=\mathop{\otimes}^{N}_{j=1}\mathop{\otimes}^{s_j}_{s=1}\rho^{s}_{i;j}$, $\rho^{s}_{i;j}$ is the reduced density matrix of the subsystems owned by the party $\textsf{A}_i$ by tracing out the subsystems owned by $\textsf{A}_j$ in arbitrary bipartite entangled pure state $|\varphi^{s}\rangle_{ij}\langle\varphi^{s}|$. The equality (\ref{eqnbipartite1}) is obtained by iteratively using the additivity of von Neumann entropy in Eq.(\ref{eqnvon}).

Moreover, the entanglement shared by two parties $\textsf{A}_i$ and $\textsf{A}_{j}$ is from entangled sources of $\mathop{\otimes}^{s_j}_{s=1}|\varphi^{s}\rangle_{ij}$. According to Eq.(\ref{eqn03}), we obtain
\begin{eqnarray}
\sum^{N}_{j=1}{\cal{E}}_{{\textsf{A}_i|\textsf{A}_j}}(\rho_{\textsf{A}_i\textsf{A}_{j}})\nonumber&=&\sum^{N}_{j=1}{\cal{E}}(\mathop{\otimes}^{s_j}_{s=1}|\varphi^{s}\rangle_{ij})
\\\nonumber&=&\sum^{N}_{j=1}S(\rho^{j}_i)
\label{eqnbipartite2}
\\&=&\sum^{N}_{j=1}\sum^{s_j}_{s=1}S(\rho^{s}_{i;j}).
\label{eqnbipartite3}
\end{eqnarray}
Similar to Eqs.(\ref{eqnbipartite0}) and (\ref{eqnbipartite1}), we obtain Eqs.(\ref{eqnbipartite3}). Thus, by combining Eqs.(\ref{eqnbipartite1}) and (\ref{eqnbipartite3}), we have
\begin{eqnarray}
{\cal{E}}_{{\textsf{A}_i|\overline{\textsf{A}_i}}}(\rho_{\textsf{A}_1\cdots\textsf{A}_n})\nonumber&=&\sum^{N}_{j=1}{\cal{E}}_{{\textsf{A}_i|\textsf{A}_j}}(\rho_{\textsf{A}_i\textsf{A}_{j}})
\\
&=&\sum^{n}_{j=1}{\cal{E}}_{{\textsf{A}_i|\textsf{A}_j}}(\rho_{\textsf{A}_i\textsf{A}_{j}}).
\label{eqnbipartite5}
\end{eqnarray}

Similar to Eq.(\ref{eqnbipartite5}), we get Eq.(\ref{eqnbipartite}) for R\'{e}nyi-$\alpha$ entropy entanglement from its definition in Eq.(\ref{eqnR1}) and the additivity of R\'{e}nyi-$\alpha$ entropy in Eq.(\ref{eqnRe}). This has completed the proof. $\Box$

\textbf{Example 4.} Consider a $4$-partite star quantum network consisting of three EPR states. It can be regarded as a $4$-partite pure state on $8\otimes2 \otimes2 \otimes2$ dimensional Hilbert space, i.e.,
\begin{eqnarray}
|\phi\rangle_{\textsf{A}\textsf{B}\textsf{C}\textsf{D}}
&=&\frac{1}{2\sqrt{2}}(|0000\rangle+|1001\rangle
\nonumber\\
&&+|2010\rangle+|3011\rangle
+|4100\rangle
\nonumber\\
&&+|5101\rangle+|6110\rangle+|7111\rangle)_{\textsf{A}\textsf{B}\textsf{C}\textsf{D}}.
\label{high0}
\end{eqnarray}
The reduced state of $\textsf{A}$ is given by
\begin{eqnarray}
\rho_{\textsf{A}}=\frac{1}{8}\sum_{i=0}^7|i\rangle\langle{}i|.
\label{re}
\end{eqnarray}
The EOF of $|\phi\rangle_{\textsf{A}|\textsf{B}\textsf{C}\textsf{D}}$ is given by: ${\cal{E}}(|\phi\rangle_{\textsf{A}|\textsf{B}\textsf{C}\textsf{D}})=3$. For mixed state $\rho_{\textsf{A}\textsf{B}}$, it should take an infimum for the EOF. Note that the reduced density matrix $\rho_{\textsf{A}\textsf{B}}$  of $|\psi\rangle_{\textsf{A}\textsf{B}\textsf{C}\textsf{D}}$ in Eq. (\ref{high0}) has a spectral decomposition
\begin{eqnarray}
\rho_{\textsf{A}\textsf{B}}\nonumber&=&\frac{1}{4}(|x_1\rangle\langle x_1|+|x_2\rangle\langle x_2|+|x_3\rangle\langle x_3|
+|x_4\rangle\langle x_4|),
\label{highAB}
\end{eqnarray}
where $|x_i\rangle$ are eigenstates given by
\begin{eqnarray}
\nonumber&&|x_1\rangle_{\textsf{A}\textsf{B}}=\frac{1}{\sqrt{2}}(|00\rangle+|41\rangle), \\\nonumber&&|x_2\rangle_{\textsf{A}\textsf{B}}=\frac{1}{\sqrt{2}}(|10\rangle+|51\rangle),
\\\nonumber&&|x_3\rangle_{\textsf{A}\textsf{B}}=\frac{1}{\sqrt{2}}(|20\rangle+|61\rangle),
\\&&|x_4\rangle_{\textsf{A}\textsf{B}}=\frac{1}{\sqrt{2}}(|30\rangle+71\rangle).
\label{}
\end{eqnarray}

By the Hughston-Jozsa-Wootters (HJW) Theorem \cite{Hughston(1993)}, any pure state ensemble of $\rho_{\textsf{A}\textsf{B}}$ can be realized as a superposition of $|x_i\rangle$ with $i=1,\cdots, 4$, that is, for arbitrary pure state $|\varphi\rangle_{\textsf{A}\textsf{B}}=\sum_ic_i|x_i\rangle$ with $\sum_ic^2_i=1$, its reduced density matrix $\rho_{\textsf{A}}={\rm Tr}_{\textsf{B}}(|\varphi\rangle_{\textsf{A}\textsf{B}}\langle\varphi|)$ has the same spectrum $\frac{1}{2}$ with double root \cite{Vidal(2002)} while other spectrum are $0$. So, we have ${\cal{E}}(|\varphi\rangle_{\textsf{A}\textsf{B}})=1$.

Moreover, the state $\rho_{\textsf{A}\textsf{B}}$ in Eq. (\ref{highAB}) can be decomposed into
\begin{eqnarray}
\rho_{\textsf{A}\textsf{B}}=\sum_ip_i|\varphi_i\rangle_{\textsf{A}\textsf{B}}\langle\varphi_i|.
\end{eqnarray}
Thus, we have
\begin{eqnarray}
{\cal{E}}(\rho_{\textsf{A}\textsf{B}})=\sum_ip_i{\cal{E}}(|\varphi_i\rangle_{\textsf{A}\textsf{B}})=1
\end{eqnarray}
from ${\cal{E}}(|\varphi_i\rangle_{\textsf{A}\textsf{B}})=1$ for each state $|\varphi_i\rangle_{\textsf{A}\textsf{B}}$.

From similar proofs, we show the EOFs of $\rho_{\textsf{A}\textsf{C}}$ and $\rho_{\textsf{A}\textsf{D}}$ are
\begin{eqnarray}
{\cal{E}}(\rho_{\textsf{A}\textsf{C}})={\cal{E}}(\rho_{\textsf{A}\textsf{D}})=1.
\end{eqnarray}
Consequently, we have
\begin{eqnarray}
{\cal{E}}(|\psi\rangle_{\textsf{A}|\textsf{B}\textsf{C}\textsf{D}})={\cal{E}}(\rho_{\textsf{A}\textsf{B}})+{\cal{E}}(\rho_{\textsf{A}\textsf{C}})+{\cal{E}}(\rho_{\textsf{A}\textsf{D}}).
\end{eqnarray}

Similarly, for the R\'{e}nyi-$\alpha$ entanglement measure in Eq.(\ref{eqnR1}), the entanglement distribution satisfies
\begin{eqnarray}
\mathcal{R}_\alpha(|\psi\rangle_{\textsf{A}|\textsf{B}\textsf{C}\textsf{D}})
=\mathcal{R}_\alpha(\rho_{\textsf{A}\textsf{B}})
+\mathcal{R}_\alpha(\rho_{\textsf{A}\textsf{C}})
+\mathcal{R}_\alpha(\rho_{\textsf{A}\textsf{D}})
\end{eqnarray}
This has completed the proof.

\subsection{Quantum network with arbitrary multipartite entangled pure states}
\label{multipartiteGHZ}

Generally, it is interesting to show the equality (\ref{eqnbipartite}) for the multipartite entangled pure state shared by two parties. Consider quantum network $\mathcal{N}_{q_{3}}$, where any two parties $\textsf{A}_i$ and $\textsf{A}_j$ share the states $\tilde{\varrho}^{1}_{ij}, \cdots, \tilde{\varrho}^{s}_{ij}$, $\sigma_{ij}(\varphi_1), \cdots, \sigma_{ij}(\varphi_t)$, and $\delta_{ij}(\vartheta_1), \cdots, \delta_{ij}(\vartheta_k)$. The joint state shared by any two parties $\textsf{A}_{i}$ and $\textsf{A}_{j}$ is given by
\begin{eqnarray}
\xi_{ij}=\mathop{\otimes}^{s_j}_{s=1}\widetilde{\varrho}^s_{ij} \mathop{\otimes}^{t_j}_{t=1} \sigma_{ij}({\varphi_t})\mathop{\otimes}^{k_j}_{k=1}\delta_{ij}(\vartheta_k),
\end{eqnarray}
where $\tilde{\varrho}^{s}_{ij}$ denotes the density matrix of the multipartite entangled pure state $|\psi\rangle_{ij}$ with bipartition, $\sigma(\varphi)$ is the density matrix of generalized GHZ state \cite{GHZ}: $|\phi(\varphi)\rangle=\cos\varphi|0\rangle^{\otimes m}+\sin\varphi|1\rangle^{\otimes m}$ with any integer $m\geq 3$, and $\delta({\vartheta})=\cos^2\vartheta|00\rangle\langle00|+\sin^2\vartheta|11\rangle\langle11|)$ is the reduced density matrix of any two subsystems obtained by tracing out other  subsystems in a generalized GHZ state. Here, $s_j$, $t_j$, and  $k_j$ denote the numbers of $\widetilde{\varrho}$, $\sigma(\varphi)$,  and $\delta(\vartheta)$ shared by two parties $\textsf{A}_i$ and  $\textsf{A}_j$.

Thus, the total state of $\mathcal{N}_{q_{3}}$ is written as
\begin{eqnarray}
\rho_{\textsf{A}_1\cdots\textsf{A}_n}=\mathop{\otimes}^{n}_{\substack{i,j=1
\\i\neq j}}\mathop{\otimes}^{s_j}_{s=1}\widetilde{\varrho}^s_{ij} \mathop{\otimes}^{t_j}_{t=1} \sigma_{ij}({\varphi_t})\mathop{\otimes}^{k_j}_{k=1}\delta_{ij}(\vartheta_k).
\end{eqnarray}
We prove that the monogamy of entanglement measure ${\cal Q}$ is valid for $\rho_{\textsf{A}_1\cdots\textsf{A}_n}$ from $\mathcal{N}_{q_3}$.

\begin{theorems}\label{bipartiteE2}
The entanglement distribution of quantum network $\mathcal{N}_{q_3}$ satisfies
\begin{eqnarray}
{\cal Q}_{{\textsf{A}_i|\overline{\textsf{A}_i}}}(\rho_{\textsf{A}_1\cdots\textsf{A}_n})\geq\sum^n_{j=1,j\neq i}{\cal Q}_{{\textsf{A}_i|\textsf{A}_j}}(\rho_{\textsf{A}_i\textsf{A}_{j}}),
\label{eqnbipartiteE2}
\end{eqnarray}
where ${\cal Q}_{{\textsf{A}_i|\overline{\textsf{A}_i}}}$ and ${\cal Q}_{{\textsf{A}_i|\textsf{A}_j}}$ are defined in Theorem \textbf{S\ref{bipartiteE}}.

\end{theorems}

\emph{Proof.} Suppose that there are $N$ parties $\textsf{A}_j$'s who are entangled with the party $\textsf{A}_i$. According to Eq.(\ref{eqn03}), by iteratively using the additivity of von Neumann entropy in Eq.(\ref{eqnRe}), we have
\begin{eqnarray}
{\cal{E}}_{{\textsf{A}_i|\overline{\textsf{A}_i}}}(\rho_{\textsf{A}_1\cdots\textsf{A}_n})
&=&
S(\rho_{\textsf{A}_i})
\nonumber\\
&=& S(\mathop{\otimes}^{N}_{j=1}(\mathop{\otimes}^{s_j}_{s=1}\rho^{s}_{i;j}\mathop{\otimes}^{t_j}_{t=1} \rho_{i;j}(\varphi_t)\mathop{\otimes}^{k_{j}}_{k=1}\rho_{i;j}(\vartheta_k)))
\nonumber
\\
&=&\sum^{N}_{j=1}S(\mathop{\otimes}^{s_j}_{s=1}\rho^{s}_{i;j}\mathop{\otimes}^{t_j}_{t=1}\rho_{i;j}(\varphi_t)
\mathop{\otimes}^{k_{j}}_{k=1}\rho_{i;j}(\vartheta_k))
\nonumber
\\
&=&\sum^{N}_{j=1}(\sum^{s_j}_{s=1}S(\rho^{s}_{i;j})+\sum^{t_j}_{t=1}S(\rho_{i;j}(\varphi_t))
\nonumber
\\
&&+\sum^{k_j}_{k=1}S(\rho_{i;j}(\vartheta_k))).
\label{eqnbipartite03}
\end{eqnarray}
Moreover, it is easy to calculate that
\begin{eqnarray}
{\cal E}_{{\textsf{A}_i|\textsf{A}_{j}}}(\rho_{\textsf{A}_i\textsf{A}_{j}})
\nonumber
&=&{\cal E}(\mathop{\otimes}^{s_j}_{s=1}\tilde{\varrho}^s_{ij} \mathop{\otimes}^{t_j}_{t=1} \sigma_{ij}({\varphi_t})\mathop{\otimes}^{k_j}_{k=1}\delta_{ij}(\vartheta_k))
\\
&\leq &
 {\cal E}(\mathop{\otimes}^{s_j}_{s=1}\tilde{\varrho}^s_{ij} )+ {\cal E}(\mathop{\otimes}^{t_j}_{t=1} \sigma_{ij}({\varphi_t}))
\label{eqnbipartite10}
\\
 &=& S(\mathop{\otimes}^{s_j}_{s=1}\rho^{s}_{i;j})+S(\mathop{\otimes}^{t_j}_{t=1} \rho_{i;j}(\varphi_t))
 \label{eqnbipartite11}
\\
 &=&\sum^{s_j}_{s=1}S(\rho^{s}_{i;j})+\sum^{t_j}_{t=1}S(\rho_{i;j}(\varphi_t)).
 \label{eqnbipartite22}
 \end{eqnarray}
From the additivity of EOF and ${\cal E}(\mathop{\otimes}^{k_j}_{k=1}\delta_{ij}(\vartheta_k))=0$, we get the inequality (\ref{eqnbipartite10}). The equality (\ref{eqnbipartite11}) follows from the definition of EOF in Eq.(\ref{eqn03}). The equality (\ref{eqnbipartite22}) is obtained by using the additivity of von Neumann entropy in Eq.(\ref{eqnvon}) iteratively. Combining Eqs. (\ref{eqnbipartite03}) and (\ref{eqnbipartite22}), it yields
\begin{eqnarray}
{\cal{E}}_{{\textsf{A}_i|\overline{\textsf{A}_i}}}(\rho_{\textsf{A}_1\cdots\textsf{A}_n})\nonumber&\geq&\sum^N_{j=1,j\neq i}{\cal{E}}_{{\textsf{A}_i|\textsf{A}_j}}(\rho_{\textsf{A}_i\textsf{A}_{j}})
\\
 &=&\sum^n_{j=1,j\neq i}{\cal{E}}_{{\textsf{A}_i|\textsf{A}_j}}(\rho_{\textsf{A}_i\textsf{A}_{j}}).
\label{eqnbipartite23}
\end{eqnarray}

Similar to the inequality (\ref{eqnbipartite23}), from Eq.(\ref{eqnR1}) and the additivity of R\'{e}nyi-$\alpha$  entropy in Eq.(\ref{eqnRe}), it is easy to get the  result in Theorem S\ref{bipartiteE2}.

\section{Extension of Theorem 1 with other entropies}
\label{entropy}

In this section, we investigate the entanglement distribution of quantum network  $\mathcal{N}_{q_2}(\cal{A},\xi)$ given in Appendix $\textsf{B}.2$ in terms of Tsallis-$q$ entanglement and Unified $(q, s)$-entropy entanglement.

\subsection{Entanglement distribution of quantum network in terms of Tsallis-$q$ entropy}
\label{Tentropy}

\begin{theorems}\label{generalT}
Assume that a quantum network  $\mathcal{N}_{q_2}(\cal{A},\xi)$ is connected by arbitrary multipartite entangled pure states.
\begin{itemize}
\item[(i)]For $0<q<1$, the entanglement distribution of $\mathcal{N}_{q_2}$ satisfies
\begin{eqnarray}
\mathcal{T}^{(q)}_{\textsf{A}_i|\overline{\textsf{A}_i}}(\rho_{\textsf{A}_1\cdots\textsf{A}_n})\geq\sum^n_{j=1,j\neq i}\mathcal{T}^{(q)}_{{\textsf{A}_i|\textsf{A}_j}}(\rho_{\textsf{A}_i\textsf{A}_{j}}).
\label{eqnT20}
\end{eqnarray}
\item[(ii)] For $q>1$, the entanglement distribution of $\mathcal{N}_{q_2}$ satisfies
\begin{eqnarray}
\mathcal{T}^{(q)}_{\textsf{A}_i|\overline{\textsf{A}_i}}(\rho_{\textsf{A}_1\cdots\textsf{A}_n})\geq\sum^n_{j=1,j\neq i}\mathcal{T}^{(q)}_{{\textsf{A}_i|\textsf{A}_j}}(\rho_{\textsf{A}_i\textsf{A}_{j}}),
\label{eqnT12}
\end{eqnarray}
\end{itemize}
where $\mathcal{T}^{(q)}_{\textsf{X}|\textsf{Y}}$ is Tsallis-$q$ entanglement in Eq.(\ref{eqnT01}) with respect to the bipartition $\textsf{X}$ and $\textsf{Y}$, and $\overline{\textsf{A}_i}$ denotes all parties except for $\textsf{A}_i$.

\end{theorems}

\emph{Proof.} Suppose that there exist $N$ parties $\textsf{A}_j$'s who are entangled with the party $\textsf{A}_i$. From Eq.(\ref{eqnT01}), for the case $0<q<1$, we have
 \begin{eqnarray}
\mathcal{T}^{(q)}_{\textsf{A}_i|\overline{\textsf{A}_i}}(\rho_{\textsf{A}_1\cdots\textsf{A}_n})\nonumber&=&S_{q}(\rho_{\textsf{A}_i})
\\
\nonumber
&=&S_{q}(\mathop{\otimes}^{N}_{j=1}(\mathop{\otimes}^{s_j}_{s=1}\rho^{s}_{i;j}))
\label{eqnT03}
\\
&\leq&
\sum^{N}_{j=1}S_{q}(\mathop{\otimes}^{s_j}_{s=1}\rho^{s}_{i;j}),
\label{eqnT30}
\end{eqnarray}
where the inequality (\ref{eqnT30}) is derived from  using the additivity of Tsallis-$q$ entropy in Eq.(\ref{eqnT02}) iteratively, $\rho^{s}_{i;j}$ is the reduced density matrix of the subsystems owned by the party $\textsf{A}_i$ by tracing out the subsystems owned by the party $\textsf{A}_j$ in a multipartite entangled pure state $|\psi^{s}\rangle$, i.e, $\rho^{s}_{i;j}={\rm{Tr}}_j(|\psi^{s}\rangle_{ij}\langle\psi^{s}|)$.

In addition, we have
\begin{eqnarray}
\sum^{N}_{j=1}\mathcal{T}^{(q)}_{\textsf{A}_i|\overline{\textsf{A}_i}}(\rho_{\textsf{A}_i\textsf{A}_{j}})
\nonumber&=&\sum^{N}_{j=1}\mathcal{T}^{(q)}(\mathop{\otimes}^{s_j}_{s=1}|\psi^{s}\rangle_{ij})
\\&=&\sum^{N}_{j=1}S_{q}(\rho^{j}_{i})
\label{eqnT13}
\\&=&\sum^{N}_{j=1}S_{q}(\mathop{\otimes}^{s_j}_{s=1}\rho^{s}_{i;j}).
\label{eqnT31}
 \end{eqnarray}
Here, the equality (\ref{eqnT13}) follows from the definition of Tsallis-$q$ entanglement in Eq.(\ref{eqnT01}), and the equality (\ref{eqnT31}) is due to $\rho^{j}_{i}=\mathop{\otimes}^{s_j}_{s=1}\rho^{s}_{i;j}$, i.e., all the states of  $\rho^{s}_{i;j}$'s are not entangled. The inequality (\ref{eqnT30}) and equality (\ref{eqnT31}) yield to the inequality (\ref{eqnT20}).

Similarly, we obtain the inequality (\ref{eqnT12}) for $q>1$. This completes the proof.

\subsection{Entanglement distribution of quantum networkd in terms of Unified $(q, s)$-entropy}
\label{Uentropy}

Similar to Theorem \textbf{S\ref{generalT}}, we obtain the following Theorem.

\begin{theorems}\label{generalU}
Assume that a quantum network $\mathcal{N}_{q_2}(\cal{A},\xi)$ consists of arbitrary multipartite entangled pure states.
 \begin{itemize}
 \item[(i)]For $0<q<1$ and $s<0$, or $q\geq1$ and $s\geq0$, the entanglement distribution of $\mathcal{N}_{q_2}$ satisfies
\begin{eqnarray}
\mathcal{U}^{(q,s)}_{{\textsf{A}_i|\overline{\textsf{A}_i}}}(\rho_{\textsf{A}_1\cdots\textsf{A}_n})\geq\sum^n_{j=1,j\neq i}\mathcal{U}^{(q,s)}_{{\textsf{A}_i|\textsf{A}_j}}(\rho_{\textsf{A}_i\textsf{A}_{j}}).
\label{eqnU11}
\end{eqnarray}
\item[(ii)] For $q>1$ and $s<0$, or $0<q<1$ and $s>0$, the entanglement distribution of $\mathcal{N}_{q_2}$ satisfies
\begin{eqnarray}
\mathcal{U}^{(q,s)}_{{\textsf{A}_i|\overline{\textsf{A}_i}}}(\rho_{\textsf{A}_1\cdots\textsf{A}_n})\geq\sum^n_{j=1,j\neq i}\mathcal{U}^{(q,s)}_{{\textsf{A}_i|\textsf{A}_j}}(\rho_{\textsf{A}_i\textsf{A}_{j}}),
\label{eqnU12}
\end{eqnarray}
\end{itemize}
where $\mathcal{U}^{(q,s)}_{\textsf{X}|\textsf{Y}}$ is Unified $(q, s)$-entropy entanglement in Eq.(\ref{eqnU01}) with respect to the bipartition $\textsf{X}$ and $\textsf{Y}$, and $\overline{\textsf{A}_i}$ denotes all parties except for $\textsf{A}_i$.

\end{theorems}

\emph{Proof.}  The proof is similar to its for Theorem \textbf{S\ref{generalT}}. Similar to the  inequality (\ref{eqnT30}), from Eq.(\ref{eqnU01}) and the subadditivity of Unified $(q, s)$-entropy for $0<q<1$ and $s<0$, or $q\geq1$ and $s\geq0$, we get
\begin{eqnarray}
\mathcal{U}^{(q,s)}_{{\textsf{A}_i|\overline{\textsf{A}_i}}}(\rho_{\textsf{A}_1\cdots\textsf{A}_n})&=&S_{q,s}(\rho_{\textsf{A}_i})
\\
\nonumber
&=&S_{q,s}(\mathop{\otimes}^{N}_{j=1}(\mathop{\otimes}^{s_j}_{s=1}\rho^{s}_{i;j}))
 \\
&\leq &\sum^{N}_{j=1}S_{q,s}(\mathop{\otimes}^{s_j}_{s=1}\rho^{s}_{i;j}).
\label{eqnU21}
\end{eqnarray}

Similar to Eq.(\ref{eqnT31}), from Eq.(\ref{eqnU01}) we have
\begin{eqnarray}
\mathcal{U}^{(q,s)}_{{\textsf{A}_i|\textsf{A}_j}}(\rho_{\textsf{A}_i\textsf{A}_{j}})=\sum^{N}_{j=1}S_{q,s}(\mathop{\otimes}^{s_j}_{s=1}\rho^{s}_{i;j}).
\label{eqnU22}
\end{eqnarray}
Thus, Eqs.(\ref{eqnU21}) and (\ref{eqnU22}) imply the inequality (\ref{eqnU11}) for $0<q<1$ and $s<0$, or $q\geq1$ and $s\geq0$.

Similarly, we can obtain the inequality (\ref{eqnU12}) for $q>1$ and $s<0$, or $0<q<1$ and $s>0$. This completes the proof.

%\section{Proof of Eq.(\ref{eqnmarginal01}) }
%\label{marginal0}
%
%Suppose that any party $\textsf{A}_i$ shares $n_1$ EPR states and $n_2$ GHZ states with the party $\textsf{A}_j$. From Eq.(\ref{eqn03}), it is clear that
%\begin{eqnarray}
%\mathcal{E}_{\textsf{A}_i|\textsf{A}_j}(\rho_{\textsf{A}_i|\textsf{A}_j})
%\nonumber
%&=&S(\rho^{j}_{\textsf{A}_i})
%\\
% \nonumber
%  &=&S((\frac{\mathbbm{1}}{2})^{\otimes n_1}\otimes (\frac{\mathbbm{1}}{2})^{\otimes n_2})
%\\
% &=&n_1S((\frac{\mathbbm{1}}{2})+n_2S(\frac{\mathbbm{1}}{2})
% \label{eqncapacity12}
%\\&=&n_1+n_2,
% \label{eqncapacity13}
%\end{eqnarray}
%where the equality (\ref{eqncapacity12}) follows from the additivity of von Neumann entropy in Eq.(\ref{eqnvon}). In Eq.(\ref{eqncapacity13}), we have used the equality of $S(\frac{\mathbbm{1}}{2})=1$.

\section{Proof of Theorem 2}

\subsection{Communication model on quantum networks}
\label{flowcut1}

In this subsection, we present necessary notions for featuring the communication on quantum networks.

Consider an $n$-partite entangled quantum network $\mathcal{N}_q({\cal{A}},\Omega)$. Let $\textbf{s}$ and $\textbf{t}$ be the source and sink, respectively. Define the \emph{flow} $f_{i \rightarrow j}$ in a quantum network $\mathcal{N}_q$ as the number of qubits which are reliably transmitted from $\textsf{A}_i$ to $\textsf{A}_j$ along the channel $\xi_{ij}$, that is,
 \begin{eqnarray}
f_{i \rightarrow j}\geq 0.
\label{E1}
\end{eqnarray}
For the quantum channel consisting of EPR states \cite{EPR} and GHZ states \cite{GHZ},
the capacity between $\textsf{A}_i$ and $\textsf{A}_j$ is defined in Eq.(\ref{eqnmarginal01}), which means the maximal number of qubits transmitted. Inspired by the method in ref.\cite{Pira(2019)}, the flow is bounded by the two-way quantum capacity of the associated channel $\xi_{ij}$, i.e.,
\begin{eqnarray}
f_{i \rightarrow j}\leq \mathcal{C}(\xi_{ij})=S(\rho^j_i),
 \label{E2}
\end{eqnarray}
where $S(\rho^j_i)$ represents the von Neumann entropy of the reduced density matrix  $\rho^j_i$ by tracing out the system of $j$, i.e., $\rho^j_i={\rm{Tr}}_j(|\xi_{ij}\rangle\langle\xi_{ij}|)$ . In addition, at any intermediate party, the number of qubits simultaneously received must be equal to the number of qubits simultaneously transmitted through all the party-to-party communications with neighbor parties. That means for any $i\in \cal{A}\setminus\{\textbf{s},\textbf{t}\}$, we have
\begin{eqnarray}
\sum_{j\in \cal{A}}f_{i \rightarrow j}=0.
 \label{E3}
\end{eqnarray}
This does not hold for the source $\textbf{s}$ and the sink $\textbf{t}$, for which we impose
\begin{eqnarray}
\sum_{j\in \cal{A}}f_{\textbf{s}\rightarrow j}=-\sum_{j\in \cal{A}}f_{\textbf{t}\rightarrow j}=f,
 \label{E4}
\end{eqnarray}
where $f$ is known as the value of the flow. This is an achievable end-to-end rate since it represents the total number of qubits which are transmitted by the source and correspondingly received by the sink via all the end-to-end communication. Thus, analogous to classical network, we define a \emph{flow} in quantum network as follows.

\begin{definition}
\label{flow}
A flow of a quantum network $\mathcal{N}_q(\mathcal{A},\Omega)$ from $\textbf{s}$ to $\textbf{t}$ is given by a function $f:\Omega\longrightarrow \mathbb{R}^{+}$, such that $f$ satisfies:
\begin{itemize}
\item[(i)] Capacity constraint: the flow is bounded by the capacity of channel, that is, for any $\xi_{ij}\in \Omega$, $f_{i\rightarrow j}$ satisfies
\begin{eqnarray}
f_{i\rightarrow j}\leq \mathcal{C}(\xi_{ij}).
\end{eqnarray}
\item[(ii)] Conservation of flow: the flow leaving the party $j$ is equal to the flow entering the  party  $j$, that is,
\begin{eqnarray}
\sum_{\{i:\xi_{ij}\in \Omega\}}f_{i \rightarrow j}=\sum_{\{k:\xi_{jk}\in \Omega\}}f_{j\rightarrow k},
\end{eqnarray}
for any party $j\in \mathcal{A}$, other than the sender(source) $\textbf{s}$ and receiver(sink) $\textbf{t}$.
\end{itemize}
\end{definition}

Herein, $f_{i \rightarrow j}$ denotes the number of qubits transmitted to from the party $i$ to $j$ by the quantum channel $\xi_{ij}$. If $f_{i \rightarrow j}=\mathcal{C}({\xi_{ij}})$, then the quantum channel $\xi_{ij}$ is saturated. Otherwise,  $\xi_{ij}$ is unsaturated.

\begin{definition}
A maximum flow $MF(\mathcal{N})$ of quantum network $\mathcal{N}_q(\mathcal{A},\Omega)$ is the maximum amount of flow $f$ from the source $\textbf{s}$ to the sink $\textbf{t}$, that is,
\begin{eqnarray}
f_{\max}=\max\{f\}.
\label{E5}
\end{eqnarray}

\end{definition}

Now, we transform the quantum network $\mathcal{N}_q(\mathcal{A}, \Omega)$ into a associated undirected graph $\mathcal{G}=({\cal V},\Omega)$, where ${\cal V}$ consists of all vertices corresponding to all the parties in $\mathcal{A}$, and $\Omega$ consists of all edges corresponding to all the entangled states in $\Omega$, in which one EPR state is schematically represented by one edge while the $m$-partite GHZ state is schematically represented by hyper edge connected by $m$ nodes \cite{Voloshin(2009)}. We then adopt the standard definition of cut-set for graph \cite{Voloshin(2009)}. A \textit{cut} $T_{cut}$ of $\mathcal{G}$ is a bipartition $(\textbf{S}_1,\textbf{S}_2)$ of ${\cal V}$ such that $\textbf{s}\in \textbf{S}_1$ and $\textbf{t}\in \textbf{S}_2$. The cut-set $\widetilde{T}_{cut}$ of $T_{cut}$ is the set of edges so that the removal of these edges disconnects the source $\textbf{s}$ and the sink $\textbf{t}$.

\begin{definition}\label{cut}
An edge cut set $\widetilde{T}_{cut}$ consists of edges such that there exists a bipartition $(\textbf{S}_1, \textbf{S}_2)$ of all the vertexes in $\cal{V}$ satisfying  $\textbf{s}\in \textbf{S}_1$ and $\textbf{t}\in \textbf{S}_2$, that is, $\widetilde{T}_{cut}=\{(i, j): i \in \textbf{S}_1, j \in \textbf{S}_2\}$. Then the capacity of an edge cut set $T_{cut}$ is given by
\begin{eqnarray}
\mathcal{C}(\textbf{S}_1,\textbf{S}_2)=\sum_{(i,j)\in\widetilde{T}_{cut}}\mathcal{C}(i, j),
 \label{eqncut1}
\end{eqnarray}
\end{definition}
where $\mathcal{C}(i,j)$ denotes the weight of edge $(i,j)$.

\begin{definition}
If $\mathcal{C}(T_{cut})\geq \mathcal{C}(T'_{cut})$ holds for any cut $T_{cut}$ in graph $\mathcal{G}$, then $T'_{cut}$ is a minimal cut of $\mathcal{G}$, that is, a minimum cut is a cut with the minimum capacity with the capacity given by
\begin{eqnarray}
\mathcal{C}(T'_{cut})=\min_{T_{cut}}\mathcal{C}(\textbf{S}_1,\textbf{S}_2).
 \label{eqncut1}
\end{eqnarray}

\end{definition}

\subsection{Proof of Theorem 2}

Firstly, we resort to following Lemma \ref{capacity0} in classical network \cite{Elias(1956),Ford(1956)}.

\begin{lemma}
\label{capacity0}
 Let $f$ be a flow on a network $\mathcal{N}$, and $T_{cut}$ be a
 cut $T_{cut}=(\textbf{S}_1, \textbf{S}_2)$ with the capacity $\mathcal{C}(T_{cut})$. Then the flow satisies
 \begin{eqnarray}
 f\leq \mathcal{C}(T_{cut}),
 \label{E8}
 \end{eqnarray}
where the equality holds if and only if each edge $(i,j)$ in the cut $T_{cut}=(\textbf{S}_1, \textbf{S}_2)$ is saturated, that is, $f_{i \rightarrow j}=\mathcal{C}(i, j)$, and each edge $(j,i)$ in $T_{cut}=(\textbf{S}_2,\textbf{S}_1)$ has zero flow with $f_{j \rightarrow i}=0$.

\end{lemma}

\emph{Proof of Lemma \ref{capacity0}}.  Denote $f=f(\textbf{S}_1, \textbf{S}_2)-f(\textbf{S}_2,\textbf{S}_1)$. On the one hand, the flow conservation law implies that $f(\textbf{S}_1, \textbf{S}_2)\leq \mathcal{C}(T_{cut})$ and  $f(\textbf{S}_2,\textbf{S}_1)\geq 0$. Note that $f=f(\textbf{S}_1, \textbf{S}_2)-f(\textbf{S}_2,\textbf{S}_1)\leq\mathcal{C}(T_{cut})$, that is,
\begin{eqnarray}
f\leq \mathcal{C}(T_{cut}).
 \label{E9}
\end{eqnarray}

For $f(\textbf{S}_1, \textbf{S}_2)\leq \mathcal{C}(T_{cut})$, the equality holds iff each edge $(i,j)$ in cut $T_{cut}=(\textbf{S}_1, \textbf{S}_2)$ is saturated. Besides,  $f(\textbf{S}_2,\textbf{S}_1)\geq 0$, where the equality holds if and only if each edge $(j,i)$ in $(\textbf{S}_2,\textbf{S}_1)$ has zero flow. So, $f=\mathcal{C}(T_{cut})$ holds if and only if each edge $(i,j)$ in the cut $T_{cut}=(\textbf{S}_1, \textbf{S}_2)$ is saturated and each edge $(j,i)$ in $(\textbf{S}_2,\textbf{S}_1)$ has zero flow. $\Box$

Now, consider a quantum network $\mathcal{N}_q({\cal{A}},\Omega)$ consists of quantum channels defined as EPR states and GHZ states. There exists a cut $T_{cut}$ in $\widetilde{T}_{cut}$ of the associated graph $\cal{G}$ separating $\textbf{S}_1$ and $\textbf{S}_2$. Similar to Lemma \ref{capacity0}, we get
\begin{eqnarray}
f(\textbf{S}_1, \textbf{S}_2)\leq \mathcal{C}(T_{cut}),
\label{E10}
\end{eqnarray}
where the equality holds if and only if each channel $\xi_{ij}$ in the cut $T_{cut}=(\textbf{S}_1,\textbf{S}_2)$ is saturated, and $f_{i \rightarrow j}=\mathcal{C}(\xi_{ij})=\mathcal{C}(i,j)$.  As its shown in refs.\cite{Bennett1,Karlsson(1998)}, one can perfectly transmit one qubit by using quantum channel of EPR state \cite{EPR} or GHZ state \cite{GHZ}. Thereby, it is feasible to achieve $f_{i \rightarrow j}=\mathcal{C}(i,j)$ from the point of view of quantum communication. Note that the transmission of information in quantum communication is  one-way, i.e., $f(\textbf{S}_2,\textbf{S}_1)=0$. Further, we get the following result.

\begin{lemma}
\label{capacity2}
If a flow $f$ and cut $T_{cut}=(\textbf{S}_1, \textbf{S}_2)$ on a quantum network $\mathcal{N}_q({\cal{A}}, \Omega)$ satisfy $f= \mathcal{C}(T_{cut})$, then $f$ is a maximum flow and the cut $T_{cut}$ is a minimum cut.

\end{lemma}

Note that any flow $f'$ of the network $\mathcal{N}_q$ satisfies $f'\leq \mathcal{C}(T_{cut})= f$. On the other hand, for any cut $T_{cut}'$ of the associated graph $\cal{G}$, it follows $\mathcal{C}'(T_{cut})\geq f=\mathcal{C}(T_{cut})$. Thus $f$ is the maximum flow and the cut $T_{cut}$ is the minimum cut. As a result, we obtain the claim in Theorem \ref{flowcut} for the quantum network $\mathcal{N}_q({\cal{A}}, \Omega)$. It means that the maximal flow of $\mathcal{N}_q({\cal{A}}, \Omega)$ is equal to the minimum-capacity cut of the associated graph $\mathcal{G}$. This completes the proof of Theorem 2.

\section{The iteration of Example 2}
\label{iteration}

In this section, embarking on presenting the example, we recall following Lemma.

\begin{lemma}
\cite{Ford(1956)}
Let $f$ be a \emph{flow} on a classical network $\mathcal{N}$. Then $f$ is the maximum flow only if $f$ has no augmenting paths.

\end{lemma}

This Lemma is also available for quantum network $\mathcal{N}_q$ consisting of EPR states and GHZ states. Hence, the main idea for the maximal flow of quantum network  $\mathcal{N}_q$ is to find augmenting paths. We apply the Ford-Fulkerson algorithm \cite{Ford(1956)} that sends the maximal possible amount of flow at each iteration, that is, the flow equals to the capacity of the path under consideration, where $f_{i}$ denotes the amount of flow sent in the iteration $i$.

\begin{figure}[h!]
\begin{center}
\resizebox{240pt}{300pt}{\includegraphics{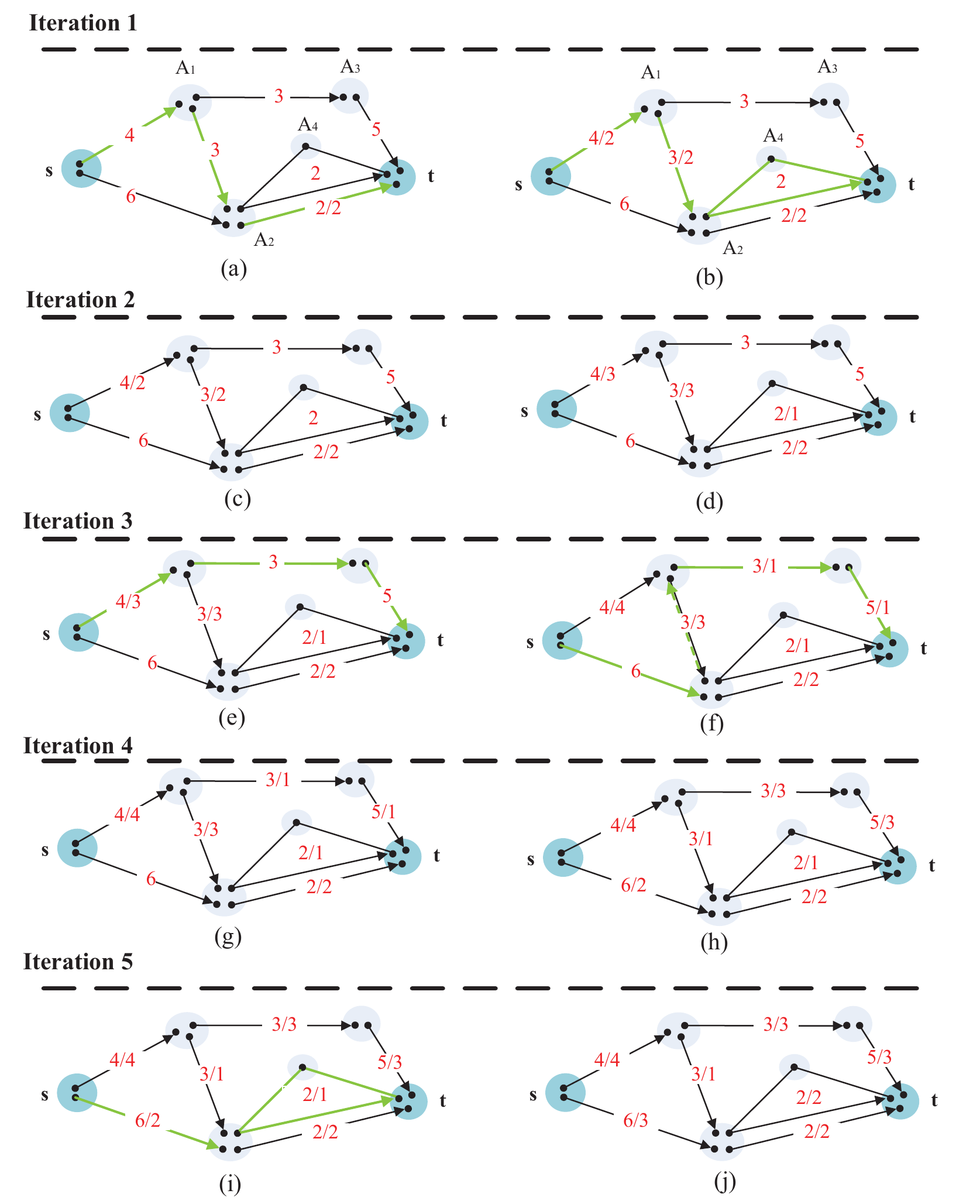}}
\end{center}
\caption{\small (Color online) The iterations for finding the maximum flow in Example 2. (a) The augmenting path $s\rightarrow 1\rightarrow 2\rightarrow t$. (b) The residual network in the first iteration by adding the maximum flow $f_{1}$. (c) The augmenting path $s\rightarrow 1\rightarrow \{2,4\}\rightarrow t$. (d) The residual network in the second iteration by adding a maximum flow $f_{2}$. (e) The augmenting path $s\rightarrow 1\rightarrow 3 \rightarrow t$. (f) The residual network in the third iteration by adding a maximum flow $f_{3}$. (g) The augmenting path $s\rightarrow 2\rightarrow 1 \rightarrow 3 \rightarrow t$. (h) The residual network in the fourth iteration by adding a maximum flow $f_{4}$. (i) The augmenting path $s\rightarrow \{2,4\} \rightarrow t$. (j) The residual network in the fifth iteration by adding a maximum flow $f_{5}$.
}
\label{flow1}
\end{figure}

\textbf{Iteration 1.} We find a path $s\rightarrow 1\rightarrow 2\rightarrow t$ that can carry a positive flow, then the maximum flow we can send along this path is $f_{1}=\min\{4,3,2\}=2$, as shown in Fig.\ref{flow1}(a). In Fig.\ref{flow1}(b), we obtain a residual network with updated link capacities resulting from pushing the flow along the path.

\textbf{Iteration 2.} Similarly, we find an augmenting path $s\rightarrow 1\rightarrow \{2,4\}\rightarrow t$ to send a positive flow. The maximum flow along this path is then given by $f_{2}=\min\{4-2,3-2,2\}=1$, as shown in Fig.\ref{flow1}(c). We get a residual network with updated capacities shown in Fig.\ref{flow1}(d).

\textbf{Iteration 3.} We find a path $s\rightarrow 1\rightarrow 3 \rightarrow t$ which sends a maximum flow along this path given by $f_{3}=\min\{4-3,3,5\}=1$, as shown in Fig.\ref{flow1}(e). A residual network is updated as shown in Fig.\ref{flow1}(f).

\textbf{Iteration 4.} For the path $s\rightarrow 2\rightarrow 1 \rightarrow 3 \rightarrow t$, a  maximum flow along this path is given by $f_{4}=\min\{6,3,3-1,5-1\}=2$ as shown in Fig.\ref{flow1}(g). A residual network with updated capacities resulting from pushing the flow along the path is given by Fig.\ref{flow1}(h).

\textbf{Iteration 5.} We find a path $s\rightarrow \{2,4\} \rightarrow t$ for sending a positive flow. The maximum flow we can send along this path is given by $f_{5}=\min\{6-2,2-1\}=1$, as shown in Fig.\ref{flow1}(i). We obtain a residual network with updated link capacities resulting from pushing the flow along the path, as shown in Fig.\ref{flow1}(j).

\section{Proof of Theorem 3}
\label{LUequivalence}

In this section, we prove Theorem \ref{equivalent} by induction $n$, that is, the number of parties in quantum network ${\cal N}_q$. Consider the following two cases.

\textbf{Case 1}. $n=3$.

\begin{figure}[htb]
\begin{center}
\resizebox{240pt}{80pt}{\includegraphics{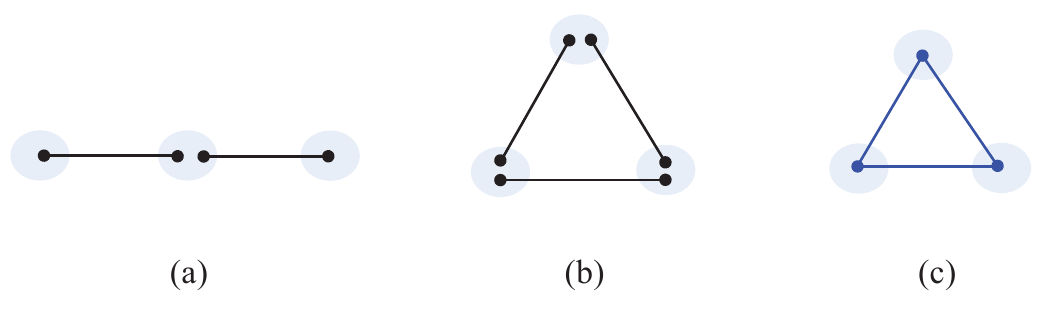}}
\end{center}
\caption{\small (Color online)  Three types of tripartite quantum network $\mathcal{N}_1$, $\mathcal{N}_2$, and $\mathcal{N}_3$ consisting of EPR states or GHZ states, where any two parties share no more than an entanglement. (a) A chain quantum network consisting of two EPR states. (b) A triangle quantum network consisting of three EPR states. (c) A triangle quantum network consisting of one GHZ state.}
\label{classify3}
\end{figure}

In this case, there are three types of quantum networks $\mathcal{N}_1$, $\mathcal{N}_2$, and $\mathcal{N}_3$, as shown in Fig. \ref{classify3}.  The vectors of von Neumann entropy of quantum networks are given by
\begin{eqnarray}
S_{\mathcal{N}_1}=(1,2,1),
\nonumber\\
S_{\mathcal{N}_2}=(2,2,2),
\nonumber\\
S_{\mathcal{N}_3}=(1,1,1).
\end{eqnarray}
It follows that the characteristic vectors of von Neumann entropy for tripartite quantum networks consisting of EPR states or GHZ states are different from each other, that is, they are unique for each kind of tripartite quantum network. Note that the von Neumann entropy is invariant under the local unitary operations. This has proved the result.

\textbf{Case 2}. $n>3$.

Assume that the result holds for $n\leq k-1$, that is, for any pair of quantum networks ${\cal N}_{1}$ and ${\cal N}_2$ there exist local unitary operations $U_j$ such that
\begin{eqnarray}
\otimes_{j=1}^nU_j\rho_{\mathcal{N}_1}U^{\dagger}_j=\rho_{\mathcal{N}_2}
\label{H1}
\end{eqnarray}
if $S_{\mathcal{N}_1}=S_{\mathcal{N}_2}$, where $S_{\mathcal{N}_j}$ denotes the characteristic vector defined by von Neumann entropies for quantum network $\mathcal{N}_j$ consisting of EPR states or GHZ states. Here, each party in networks may perform local unitary operations. It means that the quantum network consisting of EPR states and GHZ states is unique under local unitary operations.

In what follows, we will prove the result for quantum networks ${\cal N}_3$ with $n=k$. In fact, this will be proved by two subcases. The main idea is that the new network ${\cal N}_3$ can be regarded as an extended network of ${\cal N}_1$.

\begin{figure}
\begin{center}
\resizebox{200pt}{160pt}{\includegraphics{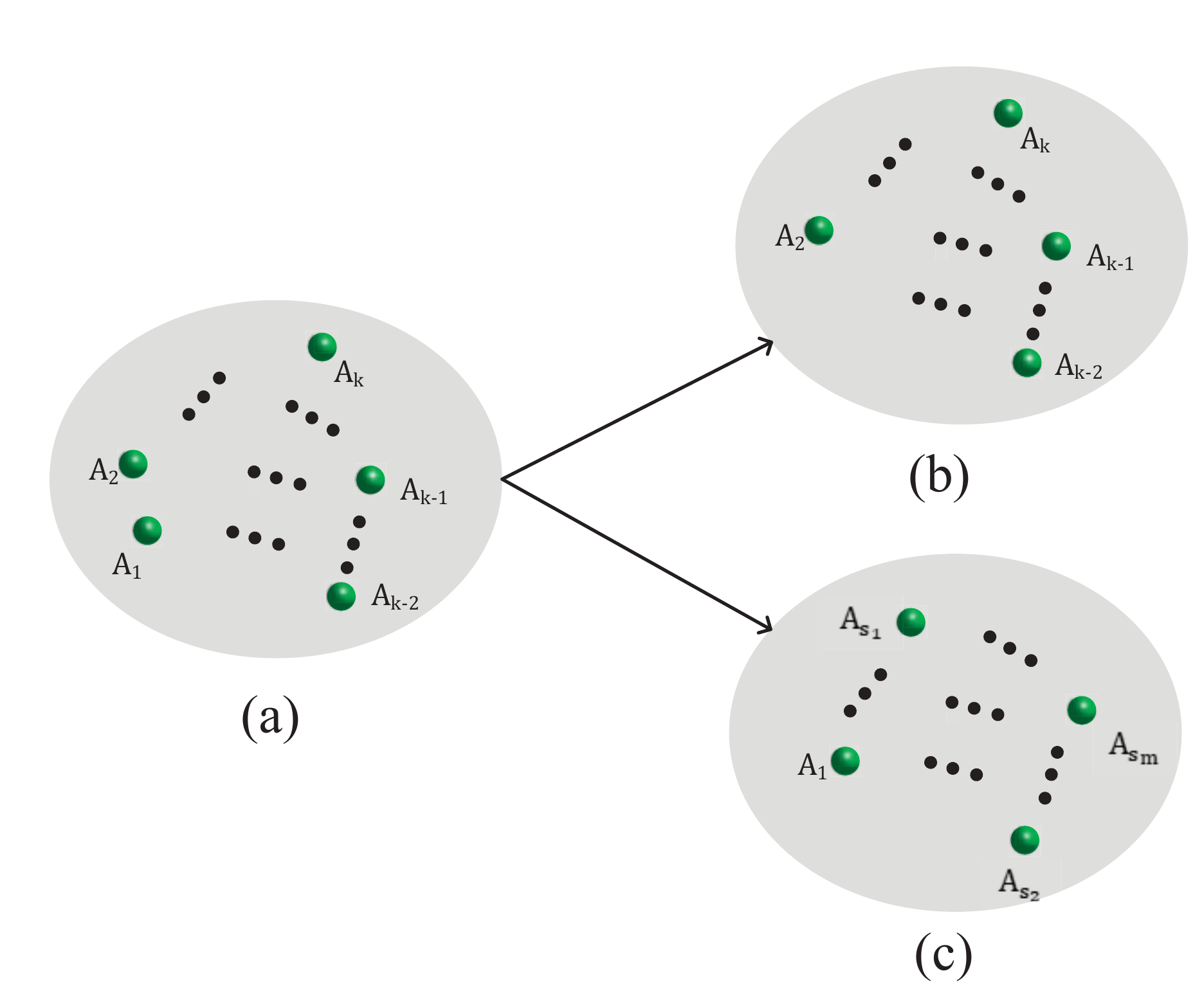}}
\end{center}
\caption{\small (Color online) Schematic quantum network ${\cal N}_3$ decomposed into two subnetworks ${\cal N}_{3;1}$ and ${\cal N}_{3;2}$. (a) ${\cal N}_3$ consisting of $\textsf{A}_1, \textsf{A}_2, \cdots, \textsf{A}_{k-1}, \textsf{A}_{k}$. (b) Subnetwork ${\cal N}_{3;1}$ consisting of $k-1$ parties ($\textsf{A}_2, \cdots, \textsf{A}_{k}$ for examples) of ${\cal N}_{1}$ and the party ${\textsf{A}}_{k}$. (c) ${\cal N}_{3;2}$  consisting of at least one party $\textsf{A}_1$ and some parties $\textsf{A}_{s_1}, \cdots, \textsf{A}_{s_m}$ entangled with $\textsf{A}_1$ of ${\cal N}_{3}$.}
\label{classify4}
\end{figure}

Assume that ${\cal N}_3$ consists of $k$ parties $\textsf{A}_1, \textsf{A}_2, \cdots, \textsf{A}_{k}$ who share EPR states and GHZ states, as shown in Fig.\ref{classify4}(a). The quantum network ${\cal N}_3$ will be decomposed into two subnetworks ${\cal N}_{3;1}$ as shown in Fig.\ref{classify4}(b) and ${\cal N}_{3;2}$ as shown in Fig.\ref{classify4}(c), where ${\cal N}_{3;1}$ consists of $k-1$ parties ($\textsf{A}_2, \cdots, \textsf{A}_{k}$ for examples) of ${\cal N}_{3}$, and ${\cal N}_{3;2}$ which has at most $k-1$ parties consists of at least one party $\textsf{A}_1$ and some parties $\textsf{A}_{s_1}, \cdots, \textsf{A}_{s_m}$ entangled with $\textsf{A}_1$ of ${\cal N}_{3}$.

Consider the subnetwork ${\cal N}_{3;1}$ with $k-1$ parties. The characteristic vector of ${\cal N}_{3;1}$ is given by $S_{{\cal N}_{3;1}}$. For the subnetwork ${\cal N}_{3;2}$, the characteristic vector is given by $S_{{\cal N}_{3;2}}$. Since these networks consist of EPR states and GHZ states. Moreover, the local von Neumann entropy of each party satisfies the additivity, namely,
\begin{eqnarray}
S_j=m^{(j)}_{EPR}+m^{(j)}_{GHZ},
\end{eqnarray}
where $m_{EPR}$ and $m_{GHZ}$ denote the respective number of EPR states and GHZ states shared by the party $\textsf{A}_j$. Note that the von Neumann entropy is invariant under the local unitary operations. It follows that the characteristic vector of quantum networks consisting of EPR and GHZ states satisfies the additivity, that is,
\begin{eqnarray}
S_{\mathcal{N}_3}=S_{\mathcal{N}_{3;1}}+S_{\mathcal{N}_{3;2}}.
\end{eqnarray}
Moreover, $\mathcal{N}_{3;1}$ and $\mathcal{N}_{3;2}$ have at most $k-1$ parties. Hence, we get that the new network generated by combinations of the subnetworks $\mathcal{N}_{3;1}$ and $\mathcal{N}_{3;2}$ is equivalent to the network $\mathcal{N}_{3}$. From the assumption, it follows that the characteristic vectors of $\mathcal{N}_{3;1}$ and $\mathcal{N}_{3;2}$ are unique under the local unitary operations. It means that for each other networks $\mathcal{N}'_{3;1}$ and $\mathcal{N}'_{3;2}$ consisting of EPR and GHZ states we have
\begin{eqnarray}
\otimes_{j}U_j\rho_{\mathcal{N}_{3;s}}U^{\dagger}_j=\rho_{\mathcal{N}'_{3;s}}, s=1,2
\label{H3}
\end{eqnarray}
if $S_{\mathcal{N}_{3;s}}=S_{\mathcal{N}'_{3;s}}$. It means the characteristic vectors of $\mathcal{N}_{3;s}$ and $\mathcal{N}'_{3;s}$ equal to each other, where $U_j$ is unitary operation. This implies that
\begin{eqnarray}
S_{\mathcal{N}_{3}}=S_{\mathcal{N}'_{3}},
\end{eqnarray}
where
\begin{eqnarray}
S_{\mathcal{N}'_3}=S_{\mathcal{N}'_{3;1}}+S_{\mathcal{N}'_{3;2}}.
\end{eqnarray}

Note Eq.(\ref{H3}) means that there exist local unitary operations satisfying
\begin{eqnarray}
\otimes_{j}U_j\rho_{\mathcal{N}_{3}}U^{\dagger}_j=\rho_{\mathcal{N}'_{3}}.
\label{H}
\end{eqnarray}
So, $\mathcal{N}_{3}$ is LU equivalent to the network $\mathcal{N}'_{3}$, that is, the characteristic vector of quantum network $\mathcal{N}_{3}$ is unique. This completes the proofs of Theorem \ref{equivalent}. $\Box$

According to Theorem \ref{equivalent}, we present a new quantum network classification method by virtue of the characteristic vector of von Neumann entropy. We present some examples to explain the main idea.

\textbf{Example 3.} Consider a simple network with $n$ parties ${\mathsf{A}_1}, {\mathsf{A}_2},\cdots, {\mathsf{A}_n}$. Here, adjacent parties share no more than one EPR-pair as shown in Fig.\ref{eqnEPRGHZ}(a) for chain quantum network, (b) for cyclic network and (c) for star quantum network. The von Neumann entropy vector of these networks are given by $(1,2,\cdots,2,1)$, $(2,\cdots,2)$, and $(n-1,1,\cdots,1)$, respectively, which are different. This shows that these networks are inequivalent to each other under local unitary operations.

\begin{figure}[htb]
\begin{center}
\resizebox{120pt}{100pt}{\includegraphics{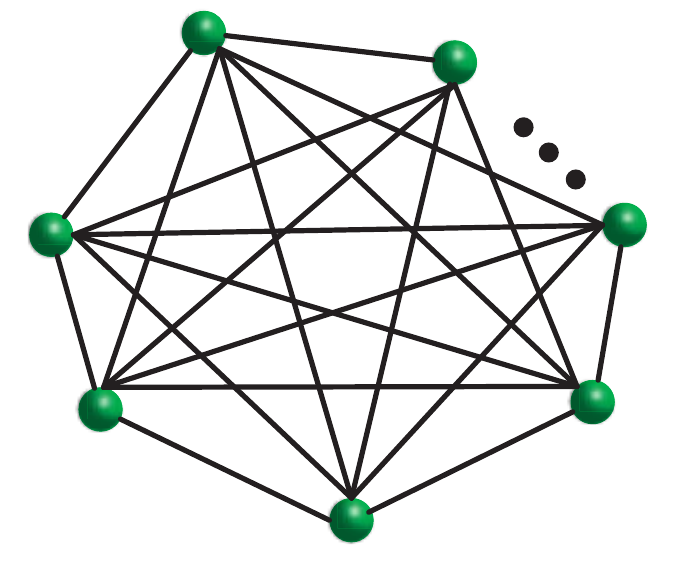}}
\end{center}
\caption{\small (Color online)  A multi-partite complete graph-like network consisting of EPR states.}
\label{completegraph}
\end{figure}

\textbf{Example 4.}  Consider an $n$-partite complete graph-like network $\mathcal{N}_\mathbf{A}$, as shown in Fig.\ref{completegraph}. Any pair of two parties shares an EPR states. Each party shares $n-1$ EPR states with others. There are $n(n-1)/2$ number of EPR states. The characteristic vector of $\mathcal{N}_\mathbf{A}$ is given by $(n-1,n-1,\cdots,n-1)$, which is different its for the networks in Fig.\ref{eqnEPRGHZ}.

\section{Classifying networks based on the mutual information}
\label{mutualinformation10}

In this section, we classify quantum networks based on the mutual information. We firstly recall the definition of mutual information.

\textbf{Shannon mutual information \cite{Shannon(1998)}} describes how much information can be determined about a random variable $\textsf{X}$, by knowing the value of a correlated random variable $\textsf{Y}$. The Shannon mutual information for tripartite system can be written as
\begin{eqnarray}
I(\textsf{X}:\textsf{Y}:\textsf{Z})\nonumber&=&H(\textsf{X}\textsf{Y}\textsf{Z})+H(\textsf{X})+H(\textsf{Y})+H(\textsf{Z})
\\&&-H(\textsf{X}\textsf{Y})-H(\textsf{Y}\textsf{Z})-H(\textsf{X}\textsf{Z}),
\label{CI}
\end{eqnarray}
where $H(\textsf{X})=-\sum_{\textsf{x}\in \textsf{X}}p(\textsf{x})\log_2p(\textsf{x})$ is the entropy of $\textsf{X}$, and $H(\textsf{X}\textsf{Y})=-\sum_{\textsf{x}\in \textsf{X},\textsf{y}\in \textsf{Y}}p(\textsf{x},\textsf{y})\log_2p(\textsf{x},\textsf{y})$ is the joint entropy of $\textsf{X}$ and $\textsf{Y}$, and $p(\textsf{x}, \textsf{y})$ is the joint probability distribution of $\textsf{X}$ and $\textsf{Y}$.

\begin{lemma}
\label{mutualinformation0}
Two types of cyclic networks in Fig.\ref{classify2}(a) and (b) can be verified by Shannon mutual information derived from local projection measurements.

\end{lemma}

\emph{Proof}. For quantum networks $\mathcal{N}_1$ and $\mathcal{N}_2$ shown in Fig.\ref{classify2}(a), they are inequivalent under local unitary operations. The total state of system for the network $\mathcal{N}_1$ is given by
\begin{eqnarray}
|\mu^1\rangle\nonumber&=&\frac{1}{\sqrt{2}}(|00\rangle+|11\rangle)_{\textsf{A}_{1}\textsf{B}_{1}}\otimes \frac{1}{\sqrt{2}}(|00\rangle+|11\rangle)_{\textsf{A}_{1}\textsf{C}_{1}}
\\&& \otimes\frac{1}{\sqrt{2}}(|00\rangle+|11\rangle)_{\textsf{B}_{1}\textsf{C}_{1}},
\label{N1}
\end{eqnarray}
which may be written into
\begin{eqnarray}
|\mu^1\rangle\nonumber&=&\frac{1}{2\sqrt{2}}(|000\rangle+|201\rangle+|012\rangle+|213\rangle
\\&&+|120\rangle+|321\rangle+|132\rangle+|333\rangle)_{\textsf{A}_{1}\textsf{B}_{1}\textsf{C}_{1}}
\label{N0}
\end{eqnarray}
where we regard the local two-qubit systems as a $4$-dimensional system with the computation basis $\{|0\rangle, \cdots, |4\rangle\}$. After projection measurements being performed on quantum state $|\mu^1\rangle$ in Eq.(\ref{N0}), we get a joint probability distribution
\begin{eqnarray}
p(x_1x_2x_3)=\frac{1}{8},
\end{eqnarray}
where $x_1x_2x_3\in \{000, 201, 012, 213, 120, 321, 132, 333\}$.

For the network $\mathcal{N}_2$ as shown in Fig.\ref{classify2}(b) the total state is given by
\begin{eqnarray}
|\mu^2\rangle=\frac{1}{\sqrt{2}}(|000\rangle+|111\rangle)_{\textsf{A}_{1}\textsf{B}_{1}\textsf{C}_{1}}^{\otimes2},
\label{N2}
\end{eqnarray}
that is,
\begin{eqnarray}
|\mu^2\rangle=\frac{1}{\sqrt{2}}(|000\rangle+|111\rangle+|222\rangle+|333\rangle)_{\textsf{A}_{1}\textsf{B}_{1}\textsf{C}_{1}}.
\label{N21}
\end{eqnarray}
For quantum state $|\mu^2\rangle$ in Eq.(\ref{N21}), after projection measurements being performed, a joint probability distribution is obtained as
\begin{eqnarray}
p(y_1y_2y_3)=\frac{1}{4},
\end{eqnarray}
where $y_1y_2y_3\in \{000,111,222,333\}$.

From Eq.(\ref{CI}), it is easy to check that the Shannon mutual information of subsystems in $\mathcal{N}_1$ and $\mathcal{N}_2$ are given by
\begin{eqnarray}
&&I(\textsf{A}_{1}:\textsf{B}_{1}:\textsf{C}_{1})=3,
\label{H9}
\\
&&I(\textsf{A}_{2}:\textsf{B}_{2}:\textsf{C}_{2})=2.
\label{H10}
\end{eqnarray}
Similarly, the Shannon mutual information of subsystems in $\mathcal{N}_3$ and $\mathcal{N}_4$ in Fig.\ref{classify2} (b) are given by
\begin{eqnarray}
&&I(\textsf{A}_{3}:\textsf{B}_{3}:\textsf{C}_{3}:\textsf{D}_{3})=4,
\\
&&I(\textsf{A}_{4}:\textsf{B}_{4}:\textsf{C}_{4}:\textsf{D}_{4})=3,
\end{eqnarray}
which are different from its in Eqs.(\ref{H9}) and (\ref{H10}). This means that the Shannon mutual information as a character can be used to distinguish two networks. $\Box$

Note that the von Neumann entropy is unitary invariant. Another interesting method for distinguishing network topologies is to use the von Neumann entropy of extended system by copying the original quantum state into two copies. Specially, by introducing an auxiliary qubit in the state $|0\rangle$ locally, each party performs a controlled Not operation on local systems, where the original qubit is controlling qubit while the auxiliary qubit is target. The total system $|\mu^1\rangle$ in Eq.(\ref{N1}) is then changed into
\begin{eqnarray}
|\nu^1\rangle
\nonumber&=&\frac{1}{\sqrt{2}}(|0\rangle^{\otimes4}+|1\rangle^{\otimes4})_{\textsf{A}_{1}\textsf{a}_{1}\textsf{B}_{1}\textsf{b}_{1}}
\\\nonumber&&\otimes\frac{1}{\sqrt{2}}(|0\rangle^{\otimes4}+|1\rangle^{\otimes4})_{\textsf{A}_{1}\textsf{a}_{1}\textsf{C}_{1}\textsf{c}_{1}}
\\&&\otimes \frac{1}{\sqrt{2}}(|0\rangle^{\otimes4}+|1\rangle^{\otimes4})_{\textsf{B}_{1}\textsf{b}_{1}\textsf{C}_{1}\textsf{c}_{1}}.
\label{N10}
\end{eqnarray}

Similarly, for the network $\mathcal{N}_2$ in Fig.\ref{classify2}(b), the state $|\mu^2\rangle$ in Eq.(\ref{N2}) is changed into
\begin{eqnarray}
|\nu^2\rangle
\nonumber&=&\frac{1}{\sqrt{2}}(|0\rangle^{\otimes6}+|1\rangle^{\otimes6})_{\textsf{A}_{2}\textsf{a}_{2}\textsf{B}_{2}\textsf{b}_{2}\textsf{C}_{2}\textsf{c}_{2}}
\\&& \otimes\frac{1}{\sqrt{2}}(|0\rangle^{\otimes6}+|1\rangle^{\otimes6})_{\textsf{A}_{2}\textsf{a}_{2}\textsf{B}_{2}\textsf{b}_{2}\textsf{C}_{2}\textsf{c}_{2}}.
\label{N20}
 \end{eqnarray}
For quantum states in Eqs.(\ref{N10}) and (\ref{N20}), we obtain
\begin{eqnarray}
&&S(\rho_{\textsf{A}_{1}\textsf{B}_{1}\textsf{C}_{1}})=3,
\\
&&S(\rho_{\textsf{A}_{2}\textsf{B}_{2}\textsf{C}_{2}})=2,
\end{eqnarray}
which are different from each other, where $S(\textsf{X})$ is the von Neumann entropy of the subsystem $\textsf{X}$. It means that we can identify two different quantum networks by using the von Neumann entropy for the system $\textsf{A}_i\textsf{B}_i\textsf{C}_i$ with $i=1,2$.

Similarly, the von Neumann entropy of the subsystems $\textsf{A}_i, \textsf{B}_i, \textsf{C}_i$ and $\textsf{D}_i$ ($i=3,4$) for $\mathcal{N}_3$ and $\mathcal{N}_4$ are respectively given by
\begin{eqnarray}
&&S(\rho_{\textsf{A}_{3}\textsf{B}_{3}\textsf{C}_{3}\textsf{D}_{3}})=4,
\\
&&S(\rho_{\textsf{A}_{4}\textsf{B}_{4}\textsf{C}_{4}\textsf{D}_{4}})=3,
\end{eqnarray}
which elucidates two inequivalent networks $\mathcal{N}_3$ and $\mathcal{N}_4$ under local unitary operations. This provides another method for classifying quantum networks.

\begin{figure}
\begin{center}
\resizebox{160pt}{300pt}{\includegraphics{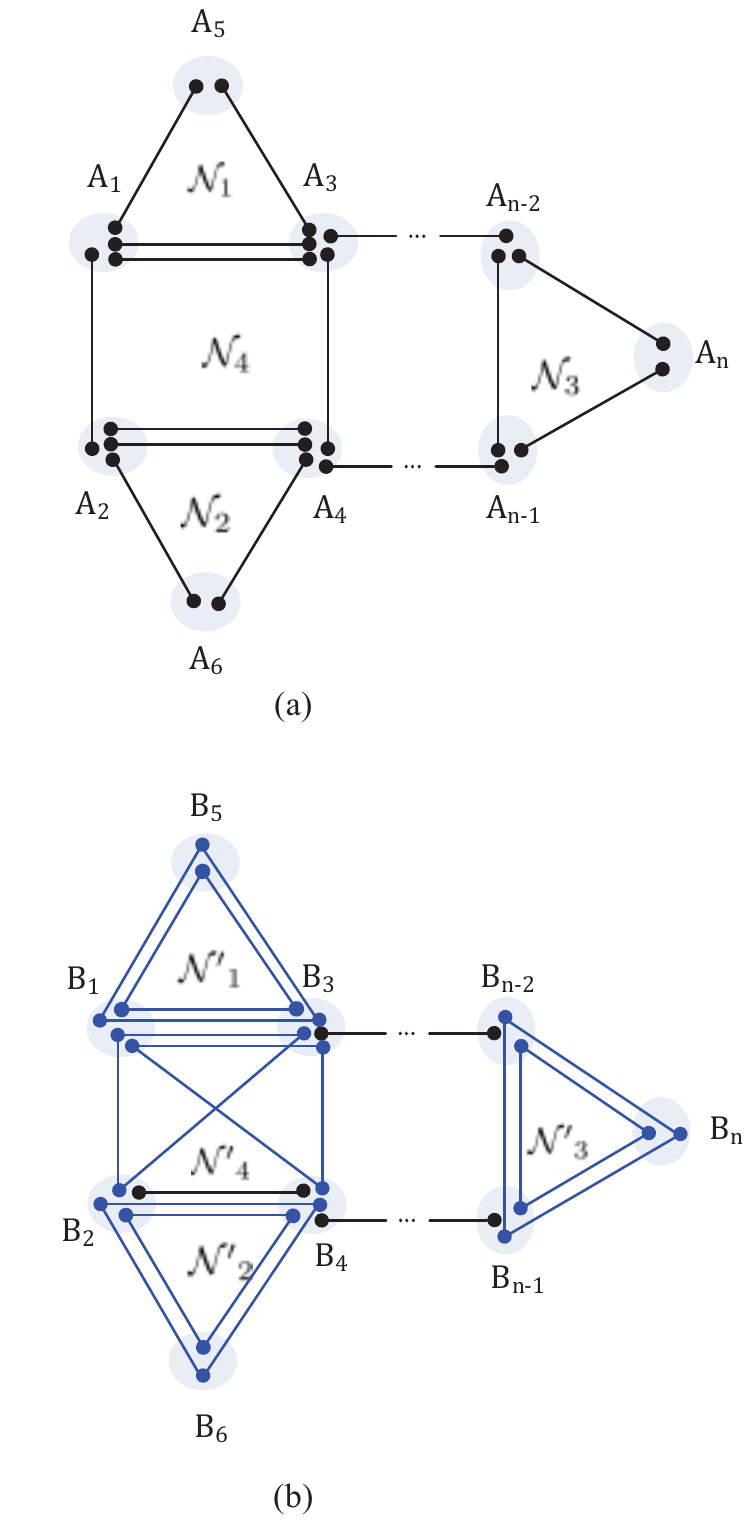}}
\end{center}
\caption{\small (Color online) Inequivalent networks with the same characteristic vector defined by von Neumann entropies. (a) Quantum network  $\mathcal{N}_{\textsf{A}}$ consisting of EPR states.(b) Quantum network $\mathcal{N}_{\textsf{B}}$ consisting of EPR states and GHZ states. }
\label{subcyclic}
\end{figure}

From Lemma \ref{mutualinformation0}, we obtain the following Theorem.

\begin{theorems}\label{mutualinformation}
For any quantum network $\mathcal{N}_{\textsf{A}}$ consisting of EPR states contains some cyclic subnetworks, there is another quantum network $\mathcal{N}_{\textsf{B}}$ with the same characteristic vector of von Neumann entropy as $\mathcal{N}_{\textsf{A}}$ such that they have different network topologies under local operations.
\end{theorems}

\emph{Proof }. As its illustrated in Fig.\ref{subcyclic}(a), quantum network $\mathcal{N}_{\textsf{A}}$ contains four subnetworks $\mathcal{N}_{1}$, $\mathcal{N}_{2}$, $\mathcal{N}_{3}$, and $\mathcal{N}_{4}$. Here,
three-partite subnetwork $\mathcal{N}_{1}$ has parties $\textsf{A}_1$, $\textsf{A}_3$ and $\textsf{A}_5$. $\mathcal{N}_{2}$ contains the parties $\textsf{A}_2$, and $\textsf{A}_4$ and $\textsf{A}_6)$. $\mathcal{N}_{3}$ contains the parties $\textsf{A}_{n-2},\textsf{A}_{n-1}$ and $\textsf{A}_{n}$. All of these subnetworks   consist of three EPR states. A four-partite cyclic subnetwork $\mathcal{N}_{4}$ consists of the parties $\textsf{A}_1$, $\textsf{A}_2$, $\textsf{A}_3$, and $\textsf{A}_4$ who share an EPR state with adjacent parties. By replacing EPR states with GHZ states, we can construct another quantum network $\mathcal{N}_{\textsf{B}}$ as shown in Fig.\ref{subcyclic}(b), where the subnetwork $\mathcal{N}_{i}$ is transformed into the corresponding subnetwork $\mathcal{N'}_{i}$, $i=1, \cdots, 4$.

From Lemma \ref{mutualinformation0}, it is easy to show that the subnetworks $\mathcal{N}_{i}$ are inequivalent to the subnetworks $\mathcal{N'}_{i}$. This follows that two networks $\mathcal{N}_{\textsf{A}}$ and $\mathcal{N}_{\textsf{B}}$ are inequivalent even if they have the same characteristic vector of von Neumann entropies as $S(\rho_{\textsf{A}_i})=S(\rho_{\textsf{B}_i})=(4,4,5,5,2,2,\cdots,3,3,2)$.

Thus, for any quantum network containing a triangle network or four-partite cyclic subnetwork consisting of EPR states, there is another quantum network with the same characteristic vector of von Neumann entropies such that they are inequivalent under local unitary operations. This completes the proof. $\Box$

Theorem S\ref{mutualinformation} shows the characteristic vector of von Neumann entropies
can not be used for classifying all quantum networks. Fortunately, this problem can be resolved by using Lemma \ref{mutualinformation0}.

\end{document}